\documentclass[aps,pra,numerical,preprint,showpacs]{revtex4-1}

\usepackage{graphicx}
\usepackage{bm}
\usepackage{amsmath}
\usepackage{amsfonts}
\usepackage{color}
\usepackage[colorlinks=true,linkcolor=blue,citecolor=blue ,urlcolor=blue]{hyperref}

\begin{document}

%\newcommand{\bm}[1]{{\bm{#1}}}
%\newcommand{\nota}[1]{\textcolor[rgb]{1,0,0}{{#1}}}

%****** new commands
\newcommand{\op}[1]{{\bm{#1}}}
\newcommand{\bra}{\langle}
\newcommand{\ket}{\rangle}
\newcommand{\new}[1]{\textcolor[rgb]{0,0,0.7}{\uline{#1}}}
\newcommand{\old}[1]{\textcolor[rgb]{1,0,0}{\sout{#1}}}
\newcommand{\nota}[1]{\textcolor[rgb]{0,0.5,0.5}{{#1}}}
%************************

\title[Dynamic generation of light states with discrete symmetries]{Dynamic Generation of Light States with Discrete Symmetries}

\author{S. Cordero} 
\email{sergio.cordero@nucleares.unam.mx}
\author{E. Nahmad-Achar}
%\email{nahmad@nucleares.unam.mx}
%
\author{O. Casta\~nos}
%\email{ocasta@nucleares.unam.mx}
%
\author{R. L\'opez-Pe\~na}
%\email{lopez@nucleares.unam.mx}
%

%
\affiliation{%
Instituto de Ciencias Nucleares, Universidad Nacional Aut\'onoma de M\'exico, Apartado Postal 70-543, 04510 M\'exico DF,   Mexico }

\begin{abstract}
A dynamic procedure is established within the generalised Tavis-Cummings model to generate light states with discrete point symmetries, given by the cyclic group ${\cal C}_n$. We consider arbitrary dipolar coupling strengths of the atoms with a one-mode electromagnetic field in a cavity. The method uses mainly the matter-field entanglement properties of the system, which can be extended to any number of $3$-level atoms. An initial state constituted by the superposition of two states with definite total excitation numbers, $\vert \psi \rangle_{M_1}$, and $\vert \psi \rangle_{M_2}$, is considered. It can be generated by the proper selection of the time-of-flight of an atom passing through the cavity. We demonstrate that the resulting Husimi function of the light is invariant under cyclic point transformations of order $n=\vert M_1-M_2\vert$. 
\end{abstract}

\maketitle

\section{Introduction}

QED in cavities is a very useful tool to explore, measure, and control quantised radiation fields and atomic systems coupled coherently within an electromagnetic resonator. Typically, a resonator is formed by two mirrors inside of which there is an atom at rest or passing through the cavity. For a single 2-level atom, the system has four parameters, viz., the spontaneous emission, the quality factor of the cavity, the atom-field coupling strength, and the travel time in the cavity. The observed effects in QED cavities include modifications in the spontaneous emission rates, changes in the atomic energy spectrum, and the continuous matter-field interchange of 
energy~\cite{walther,kimble,miller}.

Systems of $N_a$ non-interacting $2$-level atoms or molecules confined in a small container compared with the radiation wavelength have been studied within the Dicke model.  In this model, the dipolar interaction between the atoms and the field is considered~\cite{dicke}. For the one atom case, the model (called the Jaynes-Cummings model) is exactly soluble with and without the rotating wave approximation (RWA)~\cite{jaynes, zhang}. For an arbitrary number of $2$-level atoms or molecules in the RWA, the Tavis-Cummings (TC) model, has been also solved analytically under resonant conditions~\cite{cummings}. 

The Jaynes-Cummings (JC) Hamiltonian takes into account only the reversible coherent evolution, and describes a single $2$-level system strongly coupled to one mode electromagnetic field~\cite{jaynes}. The energy spectrum consists of an infinite ladder of doublets additionally to the ground state.   For a total number of excitations $M$, it is found, in resonant conditions, that the splitting of the levels is equal to $2 \sqrt{M \, \mu}$, where $\mu$ denotes the atom-field coupling strength, and this allows us to observe the anti-bunching phenomenon. Other nonlinear effects in optics have been observed in the strong coupling regime between multi-level atoms and radiation field, within the electromagnetic induced transparency (EIT) medium scheme~\cite{ciuti}. Another example in which it is necessary to consider 3-level atoms is in the area of quantum information theory, for the establishment of a quantum communication protocol in a quantum network~\cite{cirac}. 

Infinite superpositions of Fock photon states with discrete symmetries appear naturally as symmetry adapted states in the TC and Dicke models~\cite{rlp}. Since the foundations of quantum mechanics the coherent states have called the attention of the community because they minimise the Heisenberg uncertainty relations~\cite{scho}. Later works have considered superpositions of even or odd coherent states~\cite{dodonov}, and of squeezed states~\cite{walls,hollen,yuen,nieto}, all of them describing non-classical states of light because they have different statistical properties than the coherent states (which are usually called classical states of light~\cite{glauber, sud, klauder}).  The statistical properties of these infinite superpositions of Fock photon states carry irreducible representations of a finite group (called crystallised Schr\"odinger cats) have also been studied~\cite{manko}. Their evolution with respect to quadratic Hamiltonians in the quadratures of the electromagnetic field has been investigated~\cite{julio}. The proposals to generate this type of states can be grouped as follows: (i) non-linear processes~\cite{stoler,garraway}, (ii) non-demolition measurements~\cite{song,yurke2}, and (iii) field-atom interactions~\cite{banacloche,haroche,moya}. 
%%%%%%%%%%%%%

%Rydberg atoms traveling through a superconducting cavity have led to new experimental tests of quantum theory. These include the preparation of macroscopic superpositions of field states in a cavity (called Schrödinger cats)~\cite{haroche98}. The upper bounds of the relative energy difference of two Schr\"odinger cat states with a fixed fidelity have been also studied in~\cite{dodonov14}, enhancing the importance of quantum information theory.

%%%%%%%%%%%%%%%

The purpose of this work is to generate {\it finite} superpositions of photon number operator states with discrete symmetries given by the cyclic group ${\cal C}_n$ (${\cal C}_n$-states), within the generalised Tavis-Cummings model (GTC), independently of the dipolar strengths and of the number of atoms. A possible experimental setup is also presented to generate these finite superposition states with a fixed difference of the total excitation number, $\Delta M=|M_1-M_2|$, whose field part is invariant under point transformations of a cyclic group in $n$ dimensions, ${\cal C}_n$, with $n=|M_1-M_2|$.
The stationary states of the GTC model, for the one particle case, are given in analytic form together with the corresponding evolution operator. This operator is then used to study the dynamics of an arbitrary initial state by properly selecting the time-of-flight (tof) of the atom within the resonant cavity. In particular, we consider a superposition of two states with $M_1$ and $M_2$ values for the total excitation number.  The results for the one-atom case can be extended to any number of atoms, under resonant conditions as well as under detuning.

The paper is organised as follows: in section~\ref{sec2}  we discuss the generalisation of the TC model to consider atoms of 3-levels in each of the three configurations $\Xi$, $\Lambda$ and $V$. For the one-particle case, the dressed states are obtained in analytic form together with their atom-field quantum correlations measured through the calculation of the linear entropy. In section~\ref{sec3} the evolution operator for the single particle case is determined in analytic form for a given total number of excitations. This is used to study the behaviour of the light sector for two different initial conditions, which requires the reduced density matrix for the light, and from this the calculation of the Husimi function. The formation of ${\cal C}_n$-states is established with this procedure. Section \ref{sec.evolution} is dedicated to the evolution of an atom traversing a cavity, and to establish the procedure to dynamically generate ${\cal C}_n$-states of light. In section~\ref{sec5} the extension to consider any number of particles is considered. The conclusions are presented in section \ref{sec6} together with some additional remarks.

\section{Tavis-Cummings Model for $3$-level atoms}\label{sec2}

The Tavis-Cummings model describes a system of $2$-level atoms or molecules interacting dipolarly with a one-mode electromagnetic field of frequency $\Omega$, in the RWA~\cite{cummings}. This model has been used extensively to study quantum phase transitions as well as for applications in quantum information theory~\cite{lambert2004, lambert2005, physscr1-2009, physscr2-2009}.  A natural extension of the model is to consider $3$-level atoms, and the Hamiltonian in this case takes the form~\cite{eberly, cordero}
\begin{eqnarray}
{\bm H} &=&\hbar \, \Omega \, {\bm a^\dagger} {\bm a} + \hbar \, \sum^3_{j=1}\omega_{j} \, {\bm A}_{jj} - \frac{1}{\sqrt{N_a}} \sum^3_{i<j=2} \mu_{ij} ({\bm a^\dagger}  \, {\bm A}_{ij} + {\bm a} \, {\bm A}_{ji})   \, ,
\label{gtc}
\end{eqnarray}
with the convention $\omega_1 \leq \omega_2 \leq \omega_3$ for the atomic frequencies. ${\bm A}_{jk }$ denotes the raising, lowering and weight generators of the unitary algebra in 3 dimensions, which for identical particles can be realised in terms of bosonic operators as $ {\bm A}_{jk} = {\bm b}^\dagger_j \, {\bm b}_k$.  Thus, ${\bm A}_{jj}$ is the population operator for level $j$. ${\bm a^\dagger}, {\bm a}$ are the creation and annihilation photon operators, respectively.  $\mu_{jk}$ is the matter-field coupling parameter between levels $\omega_j$ and $\omega_k$, which is given by $\mu_{jk} = \hbar \,(\omega_j-\omega_k) d_{jk} \sqrt{\frac{2 \, \pi \, \rho_m}{ \hbar \, \Omega}}$, with $d_{jk}$ the matrix elements of the dipolar operator. 

\begin{figure}
\begin{center}
		\includegraphics[width=2in]{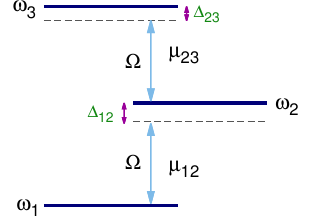}  \qquad
		\includegraphics[width=2in]{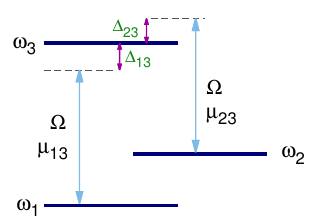}   \\[3mm]
		\includegraphics[width=2in]{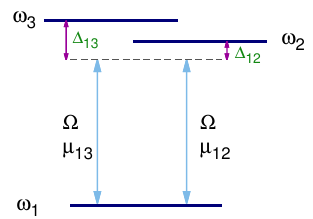}  
\end{center}
\caption{Detuning parameters $\Delta_{ij}$ for the $3$-level atomic configurations $\Xi$, $\Lambda$, and $V$, respectively (see text for details). }
\label{fig:1}       % Give a unique label
\end{figure}

In the non-resonant case we define detuning parameters $\Delta_{ij}$ in terms of the atomic frequencies $\omega_{ij}=\omega_i-\omega_j$ as (cf. Fig.~\ref{fig:1})
\begin{eqnarray*}
\Xi: &&  
\omega_{21} = \Omega + \Delta_{12}, \quad \omega_{32}= \Omega + \Delta_{23}, \quad  \omega_{31} =2 \, \Omega + (\Delta_{12} + \Delta_{23}) ; \\
V: &&
\omega_{21} = \Omega + \Delta_{12}, \quad \omega_{31}= \Omega + \Delta_{13}, \quad \omega_{32} =\Delta_{13} - \Delta_{12} \,; \\
\Lambda: &&
\omega_{31} = \Omega + \Delta_{13},  \quad \omega_{32}= \Omega - \Delta_{23}, \quad \omega_{21} =\Delta_{13} + \Delta_{23}\,.
\end{eqnarray*}

%Figura1

In the RWA the operator that measures the total number of excitations  
\begin{equation}
\bm{M}_X = \bm{a}^\dagger \, \bm{a} + \lambda_2 \, \bm{A}_{22} + \lambda_3 \, \bm{A}_{33} \, 
\label{totalE}
\end{equation}
is a constant of motion. It is dependent on the atomic configuration: the label $X$ indicates the atomic configuration $\Xi$, $V$,  and $\Lambda$, each one with $(\lambda_2, \lambda_3)=\{(1,2), (1,1), (0,1)\}$, respectively.
In particular, in this work we will be interested in the case $\bm{M}_X\geq \lambda_3 N_a$, then the dimension of the Hilbert space for all the configurations is $ \left( N_a +1 \right) (N_a+2) /2$~\cite{qts8}. This is equivalent to the Hilbert space of a three dimensional harmonic oscillator with $N_a$ quanta of energy. 

The basis states 
\begin{equation}
\vert \nu; N_a \, q \, r \rangle = \vert \nu \rangle \otimes  \sqrt{\frac{r!}{(N_a-q)! \, (q-r)! \, N_a!}} \, {\bm A}^{N_a-q}_{31} \, {\bm A}^{q-r}_{21} \vert N_a\, N_a\, N_a \rangle \, ,
\label{base}
\end{equation}
are the tensorial product of a Fock state $\vert \nu \rangle$, associated to the number of photons, and the Gelfand-Tsetlin state of $N_a$ identical particles, $\vert N_a \, q \, r\rangle$.  The lowest energy state is determined by ${\bm A}_{jk}  \vert N_a\, N_a\, N_a \rangle =0$, for all $k>1$. Here, $r$ denotes the eigenvalue of ${\bm A}_{11}$, i.e., the population of the lowest energy level $\omega_1$, and $q$ denotes the sum of the populations of the two lowest energy levels. Notice that these are eigenstates of ${\bm M}_X$ with eigenvalue $M_X= \nu + \lambda_2 \,(q-r) + \lambda_3 \,(N_a-q)$. Thus $M_X$ determines the maximum number of photons that the cavity may contain.

\subsection{One-particle case}

For $N_a=1$ all matter and field observables may be written as $3\times3$ matrices in the basis\begin{eqnarray}
&& \vert 1\rangle:= \vert M_X-\lambda_3; 1\, 0\,0\rangle, \ \ \vert 2\rangle:= \vert M_X-\lambda_2; 1\, 1\,0\rangle, \ \ \vert 3\rangle:=
\vert M_X; 1\, 1\,1\rangle,
\end{eqnarray}
where $M_X \geq \lambda_3$.  

In general, analytic expressions for the eigensystem of the Hamiltonian can be obtained for several cases, including
\begin{itemize}
\item[(i)] resonant conditions:
\begin{equation*}
\Delta_{12}=\Delta_{23}=\Delta_{13}=0 \,  \Rightarrow \omega_{31} =\lambda_3  \, \Omega  , \quad
\omega_{21} =\lambda_2 \, \Omega \, .
\end{equation*}
\item[(ii)] the following detuning conditions:
\begin{equation*}
\Xi : \Delta_{12} + \Delta_{23}=0 \, , 
\quad V : \Delta_{12} - \Delta_{13}=0 \, , \quad
\Lambda :\Delta_{13} - \Delta_{23}=0.
\end{equation*}
\end{itemize}

The obtained energy spectrum is constituted by an infinite ladder of $3$-level steps. Defining
\begin{equation}
{\cal E}_X= \sqrt{\left(\Delta_{X}/2\right)^2 + \Omega^2_X } \, ,
\end{equation}
each step is determined by 
\begin{equation}
E_{\pm} = M_X  +\Delta_X/2 \pm {\cal E}_X \, , \quad E_0= M_X \, ,
\end{equation}
with detuning values $(\Delta_\Xi,\Delta_V,\Delta_\Lambda)\equiv(\Delta_{12}, \Delta_{12}, \Delta_{13})$, and the frequencies $\Omega_X$ for the different configurations are given by  
\begin{subequations}
\begin{eqnarray}
&&  \Omega_\Xi=\sqrt{M_\Xi \, \mu^2_{12} + (M_\Xi-1) \, \mu^2_{23}}   \, , \\
&& \Omega_V= \sqrt{M_V \, \Bigl(\mu^2_{12} + \mu^2_{13}\Bigr)} \, ,  \\
&& \Omega_\Lambda= \sqrt{ M_\Lambda \, \Bigl(\mu^2_{13} +  \mu^2_{23}\Bigr)} \, . 
\end{eqnarray} 
\end{subequations}

The eigenstates of the Hamiltonian, called dressed states, can be determined in analytic form  as a linear combination of $\vert \nu \rangle \otimes \vert 1 \, q\, r\rangle \equiv \vert \nu; 1\, q\, r\rangle$. Expressions for the dressed states for the $\Xi$-, $V$- and $\Lambda$-configurations are given in~\ref{app1}.

In contrast to the bare states, which are separable, the dressed states exhibit entanglement between matter and field. It is customary to measure this entanglement through the von Neumann entropy of the reduced density matrix; for our purposes, in order to show entanglement it is sufficient to calculate the linear entropy $S_L=1-Tr{(\rho^2_F)}$, where $\rho_F$ is the reduced density matrix for the field, is different from zero. For the $\Xi$-configuration under the condition of opposite detunings $\Delta_{23}=-\Delta_{12}$, using the expressions in~\ref{app1}, we get
\begin{subequations}
\begin{eqnarray}
&& S^{(\pm)}_L  = 1 - \frac{1}{{\cal E}^4_\Xi \Bigl( 2 \pm \frac{\Delta_{12}}{{\cal E}_\Xi}\Bigl)^2}
\left\{ (M_\Xi-1)^2 \, \mu^4_{23} + M^2_\Xi \, \mu^4_{12} + \Bigl( \frac{\Delta_{12}}{2} \pm {\cal E}_\Xi \Bigr)^4 \right\}\, , \\
&& S^{(0)}_L=\frac{2 M_\Xi (M_\Xi-1) \, \mu^2_{12} \, \mu^2_{23}}{(M_\Xi \, \mu^2_{12} + (M_\Xi-1) \, \mu^2_{23})^2} \, .
\end{eqnarray}
\end{subequations}
Notice that  $S^+_L \to S^-_L$ under the change of sign of the detuning parameter. 

For the $V$-configuration, with $\Delta_{13}=\Delta_{12}$, we get $S^{(0)}_L =0$ and 
\begin{equation}
S^{(\pm)}_L  = 1 - \frac{1}{{\cal E}^4_V \Bigl( 2 \mp \frac{\Delta_{12}}{{\cal E}_V}\Bigl)^2}
\left\{ M_V^2 \, (\mu^2_{13} + \mu^2_{12})^2 + \Bigl( \mp \frac{\Delta_{12}}{2} + {\cal E}_V \Bigr)^4 \right\}  \, ,
\label{confV}
\end{equation}
while for the $\Lambda$-configuration, under the condition of equal detuning $\Delta_{23}=\Delta_{13}$, one has only to do the replacements $V \to \Lambda$, $\mu_{13} \to \mu_{23}$, and $\Delta_{12} \to -\Delta_{13}$ in expression~(\ref{confV}). Under resonant conditions both configurations yield the simplified value $S^{(\pm)}_L  = \frac{1}{2}$, independent of the total number of excitations. These expressions may also allow us to distinguish between the different configurations.

\section{Evolution operator}\label{sec3}

For the one atom case, the evolution operator associated to the Hamiltonian (\ref{gtc}) can be obtained in analytic form as $U(\tau) = e^{-i M \tau} \, U_I(\tau)$, with $\tau= \Omega \, t$, and $U_I(\tau)$ the evolution operator in the interaction picture, given by
\begin{eqnarray}
\bm{U}_{I}(\tau)&&=
U_{11}(\tau) \,\vert M-\lambda_{3};\,100\rangle\langle M-\lambda_{3};\,100\vert
+U_{12}(\tau) \,\vert M-\lambda_{3};\,100\rangle\langle M-\lambda_{2};\,110\vert
\nonumber\\
&&+U_{13}(\tau) \,\vert M-\lambda_{3};\,100\rangle\langle M;\,111\vert
\nonumber\\
&&+U_{21}(\tau) \,\vert M-\lambda_{2};\,110\rangle\langle M-\lambda_{3};\,100\vert
+U_{ 22}(\tau) \,\vert M-\lambda_{2};\,110\rangle\langle M-\lambda_{2};\,110\vert
\nonumber\\
&&+U_{ 23}(\tau) \,\vert M-\lambda_{2};\,110\rangle\langle M;\,111\vert
\nonumber\\
&&+U_{ 31}(\tau) \,\vert M;\,111\rangle\langle M-\lambda_{3};\,100\vert
+U_{ 32}(\tau) \,\vert M;\,111\rangle\langle M-\lambda_{2};\,110\vert
\nonumber\\
&&+U_{ 33}(\tau) \,\vert M;\,111\rangle\langle M;\,111\vert
\end{eqnarray}
The explicit form of the evolution operator depends of the considered atomic configuration through $\lambda_2$ and $\lambda_3$, and the corresponding matrix elements are given in~\ref{app2}.

Given the experimental and technological advances, it is possible to prepare a resonant cavity with a definite number of photons~\cite{walther}. It is then relevant to consider the evolution of an atom in its ground state in a QED cavity with $\nu_0$ photons; at time $\tau$ the state is given by the expression
\begin{equation}
\vert \phi_{\nu_0}(\tau) \rangle = U_{13}(\tau) \vert \nu_0- \lambda_3; 100 \rangle + U_{23}(\tau) \vert \nu_0- \lambda_2; 110 \rangle + U_{33}(\tau) \vert \nu_0; 111 \rangle \, .
\label{evo1}
\end{equation}
From this state we construct the corresponding density matrix and, by taking the partial trace with respect to the matter sector, one gets the reduced density matrix for the electromagnetic field 
\begin{eqnarray}
\rho_F(\nu_0,\tau)&=&  \vert U_{13}(\tau)\vert^2 \vert \nu_0 - \lambda_3\rangle\langle \nu_0- \lambda_3\vert  + \vert U_{23}(\tau)\vert^2 \vert \nu_0 - \lambda_2\rangle\langle \nu_0- \lambda_2\vert  \nonumber \\
&& + \vert U_{33}(\tau)\vert^2 \vert \nu_0 \rangle\langle \nu_0 \vert \, .
\label{red}
\end{eqnarray}
The probability to have $\nu$ photons in the cavity at time $\tau$ then takes the form
\begin{equation}
{\cal P}(\nu,\tau)=  \vert U_{13}(\tau)\vert^2 \delta_{\nu, \nu_0 -\lambda_3} + \vert U_{23}(\tau)\vert^2 \delta_{\nu, \nu_0 -\lambda_2} + \vert U_{33}(\tau)\vert^2 \delta_{\nu, \nu_0 }  \, .
\end{equation}

This yields a Husimi function for the state of light which is a linear combination of Poissonian distributions, weighted by the corresponding probability to find $\nu$ photons.
An animation of the evolution of the state of light inside the cavity, as given by this Husimi function, for a total excitation number $M$, is given in the {\it Supplementary Material} for the cases $\nu_0=2$ and $\nu_0=7$ and an atom in the $\Xi$-configuration.

\subsection{Husimi function: Electromagnetic field }

If we consider a superposition of matter-field states of the form
\begin{equation}
\vert \Phi (0) \rangle = \cos\theta \vert \nu_1; 1\,1\, 1\rangle + e^{i \xi} \sin\theta \, \vert \nu_2; 1\,1\,1\rangle \, ,
\label{initial}
\end{equation}
with $\nu_1 \neq \nu_2$, the reduced density matrix can be obtained as before and the Husimi function is calculated by taking the expectation value with respect to the Glauber coherent state of light $\vert \alpha \rangle$:
\begin{eqnarray}
 \hspace{0.5cm} Q_H(\varrho,\phi, \theta,\xi) &=& \frac{e^{-\varrho^2}}{\pi}   \Bigl( \cos^2\theta \, \frac{\varrho^{2 \, \nu_1}}{\nu_1!} + \sin^2\theta \, \frac{\varrho^{2 \, \nu_2}}{\nu_2!}  \nonumber \\
&& \hspace{1.2cm}+ \varrho^{\nu_1+\nu_2} \, \frac{\sin2 \, \theta}{\sqrt{\nu_1! \, \nu_2!} } \cos{ \left[(\nu_1-\nu_2) \phi - \xi\right]}\Bigr) \, , 
\end{eqnarray}
where $\alpha=\varrho \, e^{i \phi}$. This quasi-distribution probability function is invariant under the transformation $\phi \to \phi + 2 \pi/(\nu_1-\nu_2)$. That is, the Husimi function exhibits a cyclic point symmetry, $C_{\vert \nu_1-\nu_2\vert}$. Notice that, at this stage, this result can be extended to any number of particles because we have only to replace the matter sector of the state $\vert 1\, 1\, 1\rangle \to \vert N_a\, N_a \,N_a \rangle$, which disappears after the tracing operation.

The dynamics of state~(\ref{initial}) can be calculated through the action of the unitary evolution operator, in other words,
\begin{equation}
\vert \Phi(\tau) \rangle = \cos\theta \, \vert \phi_{\nu_1} (\tau) \rangle + e^{i \xi} \sin\theta \, \vert \phi_{\nu_2} (\tau) \rangle \, , 
\end{equation}
with $\vert \phi_{\nu_{a}} (\tau) \rangle$, $a=1,2$ given by Eq.~(\ref{evo1}).  Calculating its density operator and taking the partial trace with respect to the matter sector, the reduced density matrix for the electromagnetic field takes the form  
\begin{eqnarray}
 \rho_F(\nu_1,\nu_2,\tau) &=& \cos^2 \theta \, \rho_F(\nu_1,\tau) + \sin^2 \theta \, \rho_F(\nu_2,\tau) \nonumber \\
&& + \sin\theta \, \cos\theta \,\, \hbox{Tr}_{\textsc{m}}\left(  e^{-i \xi} \ \vert \phi_{\nu_1} (\tau) \rangle\langle \phi_{\nu_2}(\tau) \vert  +  e^{i \xi} \ \vert \phi_{\nu_2} (\tau) \rangle\langle \phi_{\nu_1}(\tau) \vert \right) \, ,
\label{rhof}
\end{eqnarray}
where $\rho_F(\nu,\tau)$ is obtained from expression~(\ref{red}), and
\begin{eqnarray}
\hbox{Tr}_{\textsc{m}}(\ldots) =&& \vert U_{13}(\tau)\vert^2\, \left[ e^{-i\xi}\vert\nu_1 - \lambda_3\rangle \langle\nu_2-\lambda_3\vert + h.c. \right] \nonumber\\
&&+ \vert U_{23}(\tau)\vert^2\, \left[ e^{-i\xi}\vert\nu_1 - \lambda_2\rangle \langle\nu_2-\lambda_2\vert + h.c. \right] \nonumber\\
&&+ \vert U_{33}(\tau)\vert^2\, \left[ e^{-i\xi}\vert\nu_1\rangle \langle\nu_2\vert + h.c. \right]\, .
\end{eqnarray}

From this expression one calculates the Husimi function by taking the expectation value with respect to the coherent states
\begin{equation}
Q_H(\nu_1,\nu_2,\alpha,\theta,\xi,\tau) = \frac{1}{\pi} \, \langle \alpha \vert \rho_F(\nu_1,\nu_2,\tau)\vert \alpha \rangle \, , 
\label{qh}
\end{equation}
which again is invariant under point transformations of the cyclic group ${\cal C}_{\vert \nu_1- \nu_2\vert}$. The extension of this result to any number of particles is not straightforward. However, as we will see below, this result is maintained.

\subsection{$\Xi$-Configuration}
%Figura 2
\begin{figure}
\begin{center}
		\includegraphics[width=0.3\linewidth]{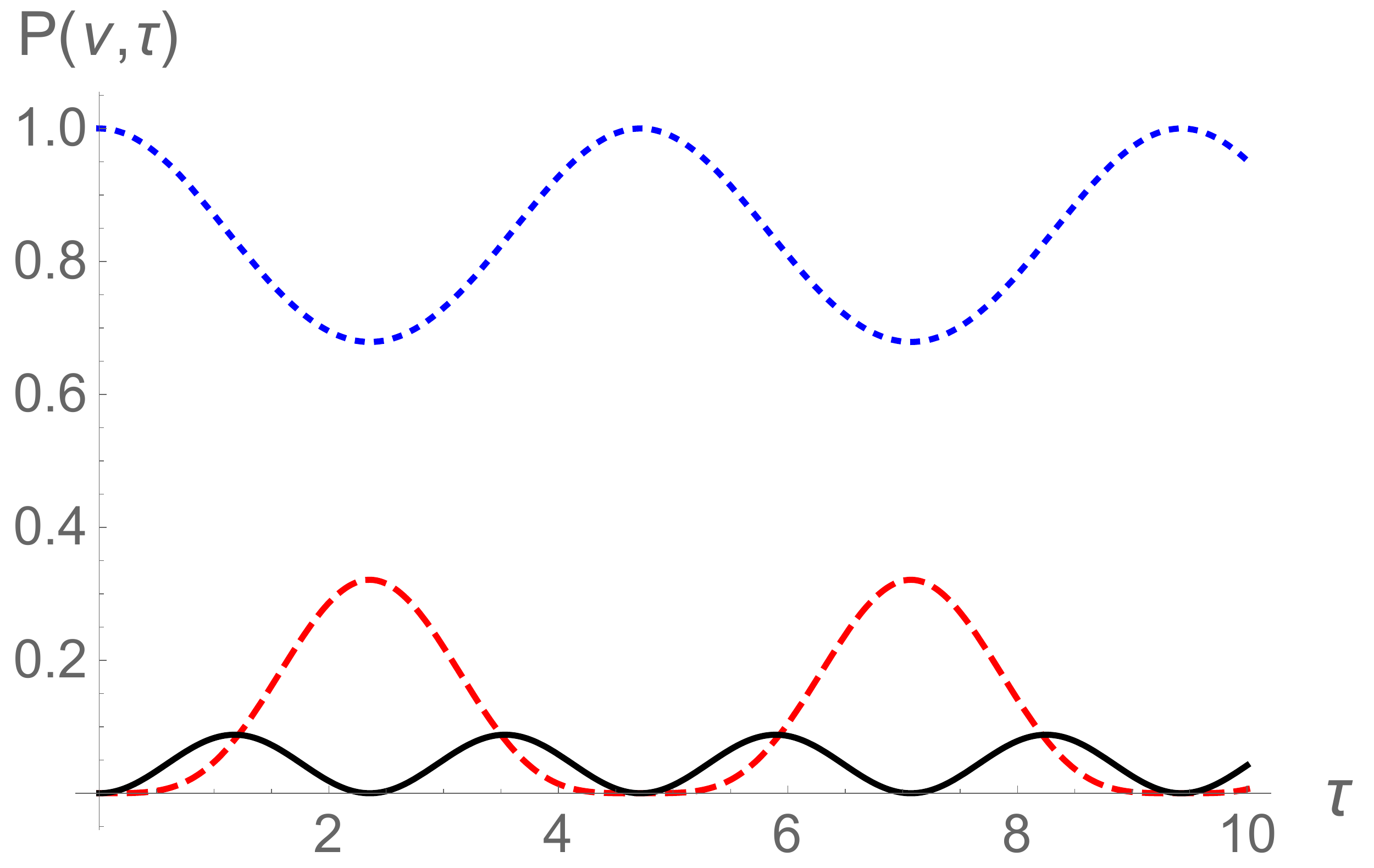}  \quad
		\includegraphics[width= 0.3\linewidth]{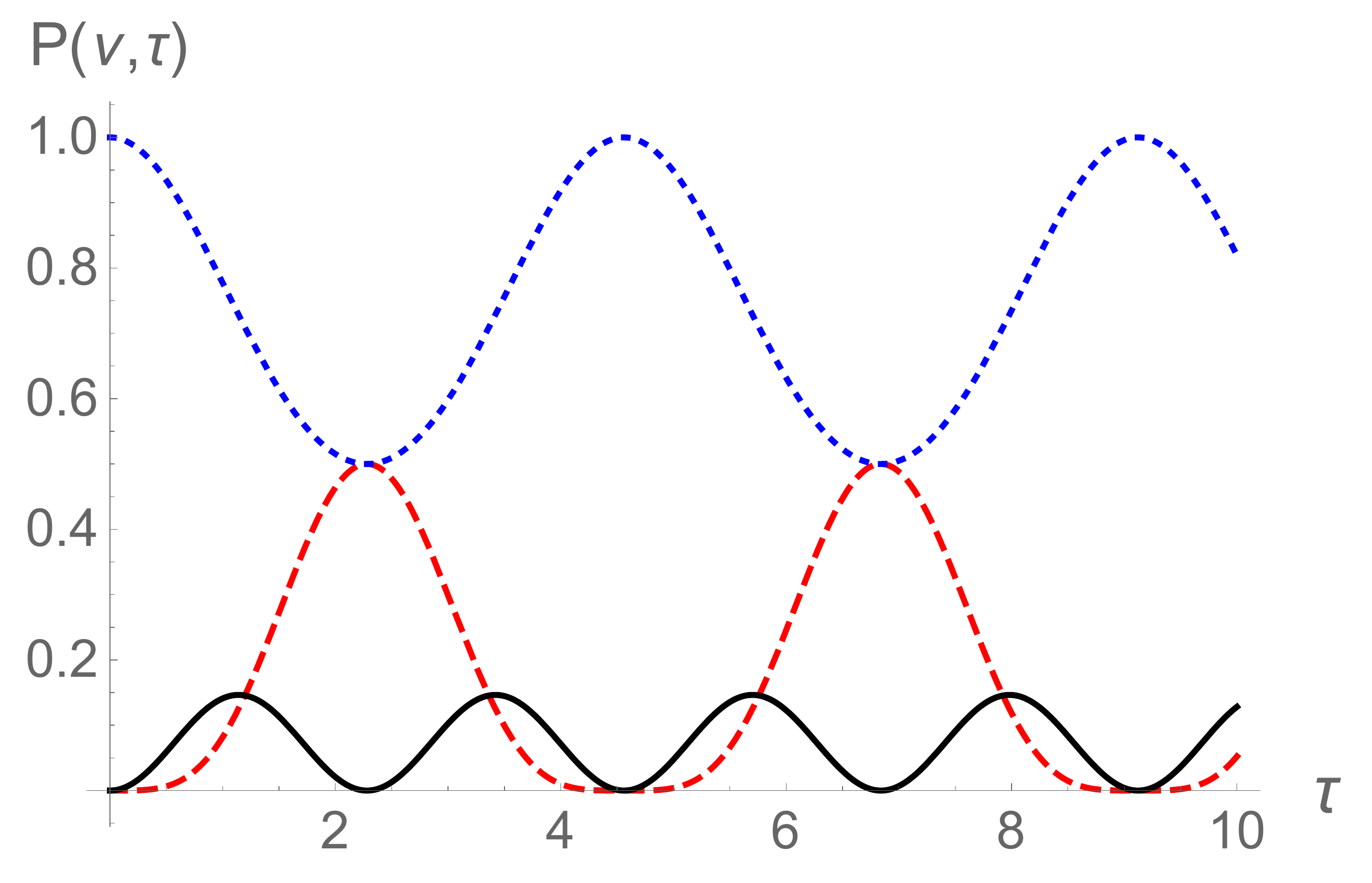}  \quad
		\includegraphics[width= 0.3\linewidth]{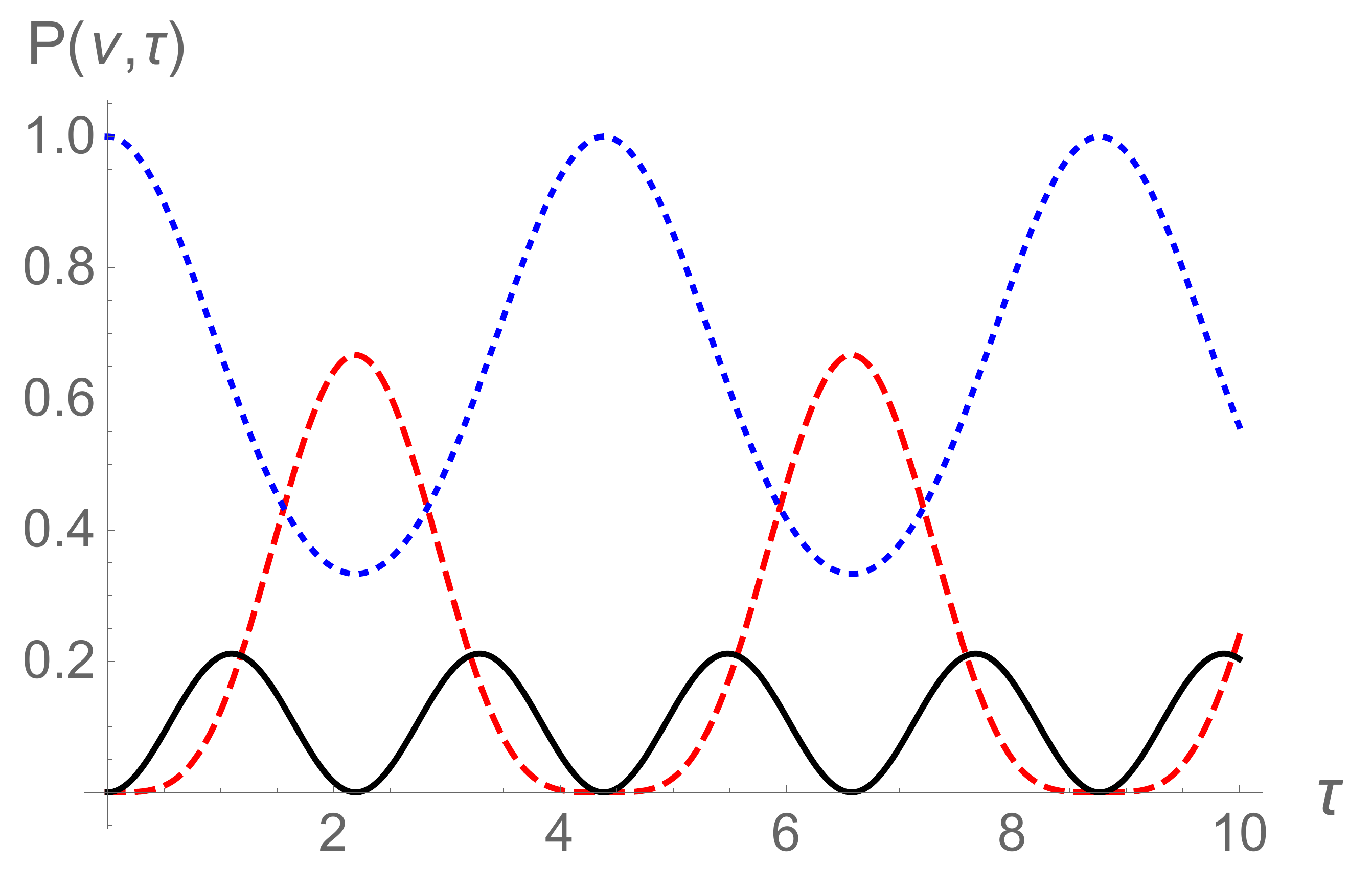}  
\end{center}
\caption{The probability to find $\nu-2$ (red, dashed line), $\nu-1$ (black, solid line) and $\nu$ (blue, dotted line) photons, in the state $\vert \psi_{\nu_0}(\tau) \rangle$, as a function of dimensionless time $\tau= \Omega \, t$, for the $\Xi$-configuration.  From left to right: $\mu_{12}< \mu^{(s-)}_{12}$, $\mu_{12} = \mu^{(s-)}_{12}$, and $\mu_{12}> \mu^{(s-)}_{12}$.}
\label{pphoton}
\end{figure}
Here we apply the previous results for a $3$-level atom in the ladder configuration. We start by calculating 
the photon number probabilities, which are shown in Fig.~\ref{pphoton}. Since the probability amplitude of finding $\nu_0 - 1$ photons is given by 
\begin{equation}
U_I(\tau)_{23}= i \sqrt{M_\Xi} \ \mu_{12} \frac{\sin{{\cal E}_\Xi \tau}}{{\cal E}_\Xi} \, e^{-i \Delta_{12} \, \tau/2} \, , 
\end{equation}
we can always make it vanish by appropriately choosing the time 
\begin{equation}
t_s = n\,\pi/{\cal E}_\Xi,\quad n\in\mathbb{N} \, .
\label{times}
\end{equation}
We may then have a resonant cavity in a superposition of states with a difference of two photons, that is, with ${\cal P}(\nu_0-1,t_s)=0$. Furthermore, by requiring 
\[
{\cal P}(\nu_0-2, t_s)={\cal P}(\nu_0, t_s)
\]
we have
\begin{eqnarray}
\hspace{1cm}
\cos\left(\frac{\Delta_{12}\,\pi}{\sqrt{4 \nu_0 (\mu_{12}^2 + \mu_{23}^2) + \Delta_{12}^2-4\mu_{23}^2}}\right)= \frac{(\nu_0\,\mu_{12}^2-(\nu_0-1)\mu_{23}^2)^2}{4\nu_0 (\nu_0-1) \mu_{12}^2\,\mu_{23}^2}\,,
\end{eqnarray}
where $\Delta_{12}$ is still a free parameter.
When in resonance, $\Delta_{12}=0$,
\begin{eqnarray}
\mu^{(s\pm)}_{12} &=& (\sqrt{2}\pm1)\ \sqrt{\frac{\nu_0-1}{\nu_0}}\,\mu_{23}\ . 
\end{eqnarray}
This result will play an important role in Section~\ref{sec.evolution}.

Note that it is not necessary to require ${\cal P}(\nu_0-2, t_s)={\cal P}(\nu_0, t_s)$; having them different from zero is enough, and this may be achieved for any value of $\mu_{12}$ as shown in Fig.~\ref{pphoton}, allowing us to write any linear superposition of these two-photon states as
\[
\vert\psi(0)\rangle_{\rm F}= \cos\theta \vert \nu_0-2\rangle + e^{i \xi} \sin\theta \vert \nu_0\rangle \, .
\]
We will establish below a robust process for building such a superposition state.

\section{Evolution of one atom traversing a cavity}
\label{sec.evolution}

In this section, we show how to generate dynamically a ${\cal C}_n$-state of light by using resonant or near resonant atoms with the mode field of the cavity. A possible experimental setup could be the following: A resonant cavity is prepared to have a definite number of photons, $\nu=\nu_0$, with $\nu_0>2$~\cite{haroche, walther}. Then a $3$-level atom in the $\Xi$-configuration, in the ground state is sent through the cavity.  Correlations between matter and field are established, which allow us to select the exit time $t_s$ of the atom to leave the cavity with an electromagnetic state in a superposition with $\nu_0-2$ and $\nu_0$ photons, approximately with the same probability, i.e., $t_s$ is chosen so that the amplitude of the contribution with $\nu_0-1$ photons vanishes (cf. Fig.~\ref{pphoton}). The cavity is left in a state with a Husimi function which has a ${\cal C}_2$-symmetry.  Immediately, a second atom, in its lower level, is sent through the cavity. As it enters, it feels a time-dependent matter-field coupling forming a superposition of the following form as initial state:
\begin{equation}
\label{init.state}
\vert\psi(0)\rangle= \left(\cos\theta \vert \nu_0-2\rangle + e^{i \xi} \sin\theta \vert \nu_0\rangle\right)\otimes\,
\vert 1\,1\,1\rangle \, ,
\end{equation}
where $\xi$ is an additional phase. This state will evolve while the atom traverses the cavity, until it leaves. As, we have shown in the previous section, the Husimi function which describes the light maintains a ${\cal C}_2$ point symmetry at all times, that is, the state in the cavity is a ${\cal C}_2$-state.

So much for the mental picture. We now establish the Schr\"odinger equation with 
time-dependent matter-field coupling interaction strengths. The time-dependent coupling $\mu(\tau)$ for a time-of-flight $t=t_{\rm tof}$ (in units of the frequency of the field) through the cavity may be described as
\begin{widetext}
\begin{equation*}
\mu(t_{\rm tof}, t) = \left\{
\begin{array}{cl}
   \frac{\exp[- \frac{t_{\rm tof}}{t (t_{\rm tof}-t)}]}
   {\left(\exp[-\frac{1}{t}]\, +\, \exp[-\frac{1}{1-t}] \right)\, \left(\exp[-\frac{1}{t_{\rm tof}-t}]\, +\, \exp[-\frac{1}{1-t_{\rm tof}+t}] \right)} &  {\rm for} \quad  0<t<t_{\rm tof}  \\[7mm]
     0 &    {\rm otherwise}
\end{array}
\right. \, .
\end{equation*}

The explicit expressions for $\mu_{ij}$ will be given in terms of the previous function, which is differentiable to all orders because it is a partition of unity.

For an arbitrary initial state $|\psi(0)\rangle = |\psi_0\rangle$ one may write its evolution in terms of the states $| \nu; \,1\, q\,r\rangle$. The matrix elements of the different atomic operators can be calculated~\cite{cordero}. Thus the time-dependent solution to the Schr\"odinger equation with initial condition $|\psi(0)\rangle = |\psi_0\rangle$ may be written as
\begin{equation}\label{eq.psi.t}
|\psi(\tau)\rangle_M = \sum_{q=0}^{1}\sum_{r=0}^q \phi_{M+q+r-2, qr}(\tau)\, e^{-i \,{\cal E}_{M+q +r-2, \, q\, r} \tau}\,|M+q +r-2;\,1\,q\,r\rangle ,
\end{equation}
\end{widetext}
where  $\phi_{M+q+r-2, qr}(\tau)$ is a time-dependent coefficient which is determined by considering the coupling interaction of the Hamiltonian and
\begin{equation}\label{eq.Enuqr}
 {\cal E}_{M+q +r-2, q \, r} = \Omega\, M + (1-q)\, \Delta_{23} + (1 - r) \, \Delta_{12} \, ,
\end{equation}
is the energy given by the diagonal contribution of the Hamiltonian for each state $|\nu;\,1\,q\,r\rangle$, i.e., $\bm{H}_D\,|\nu;\,1\,q\,r\rangle={\cal E}_{\nu qr}\,|\nu;\,1\,q\,r\rangle$. Notice that we are {\it not } necessarily considering the resonant case. For the resonant case $\Delta_{23}=\Delta_{12}=0$ 
one has that ${\cal E}_{M+q +r-2, q \, r}= \Omega \, M$.
Since $|\psi(\tau)\rangle$ should be  a solution of the time-dependent Schr\"odinger equation, the coefficients $\phi_{\nu qr}(\tau)$ must obey the following system of coupled differential equations
%\begin{widetext}
\begin{eqnarray}\label{eq.dphi}
&& \dot\phi_{M+q+r-2, \, q \, r }(\tau) = -i  \sum_{q'=0}^{1}\sum_{r'=0}^{q'}\, e^{-i({\cal E}_{M+q'+r'-2,\,q'r'}-{\cal E}_{M+q+r-2, \, q \, r})\tau}\,\nonumber\\
&&\qquad\times\langle M+q+r-2;\,1\, q\, r  \vert \bm{H}_{int}\vert M+q'+r'-2;\,1 \, q' \, r' \rangle \, \phi_{M+q'+r'-2, \, q' \, r'}(\tau),
\end{eqnarray}
where the matrix elements of $\bm{H}_{int}\equiv - \frac{1}{\sqrt{N_a}} \sum^3_{i<j=1} \mu_{ij} \,\left(\bm{a}^\dag \, \bm{A}_{ij}+ \bm{a} \, \bm{A}_{ji}\right)
$ are given by
\begin{eqnarray}
\label{mat.elem}
&&  \hspace{0.5cm} \langle\nu';\,1 \,q' \,r'  \vert \bm{H}_{int}\vert \nu;\,1 \,q\,r\rangle\nonumber \\
&&=-\mu_{12}(\tau) \left(\sqrt{(\nu+1)(q-r)(r+1)} \, \delta_{\nu' \nu+1} \, \delta_{r'r+1} 
+ \sqrt{\nu(q-r+1)r}\,\delta_{\nu' \nu-1}\delta_{r'r-1} \right)\,\delta_{q'q} \nonumber \\[2mm]
&& -\mu_{13}(\tau)\left(\sqrt{(\nu+1)(1-q)(r+1)}\,\delta_{\nu' \nu+1}\delta_{q'q+1}\delta_{r'r+1} \right. \nonumber \\
&& \left. \hspace{2cm} +\sqrt{\nu\,(2-q)\,r}\,\delta_{\nu' \nu-1}\delta_{q'q-1}\delta_{r'r-1}\right)\nonumber \\[2mm]
&&   - \mu_{23}(\tau)\left(\sqrt{(\nu+1)(1-q)(q-r+1)}\,\delta_{\nu'\nu+1}\delta_{q'q+1}\delta_{r'r} \right. \nonumber \\
&& \left. \hspace{2cm} +\sqrt{\nu\,(2-q)\,(q-r)}\,\delta_{\nu'\nu-1}\delta_{q'q-1}\delta_{r'r}  \right)\,.
\end{eqnarray}
The system of coupled differential equations for $\phi_{\nu q r}(\tau)$ (\ref{eq.dphi}) can be rewritten in matrix form:
\begin{equation}
\label{sys.eqs}
\frac{d}{d t} \, \bm{\phi} = \bm{W}(\tau) \, \bm{\phi} \, ,
\end{equation}
where $\bm{W}$ is a matrix with elements $W_{\kappa_1,\kappa_2}(\tau)$, for indices $\kappa_1,\kappa_2=1,2,3$, since we are interested in considering cases where $M \geq 2$.

\begin{figure}
\begin{center}
		\includegraphics[width=0.45\linewidth]{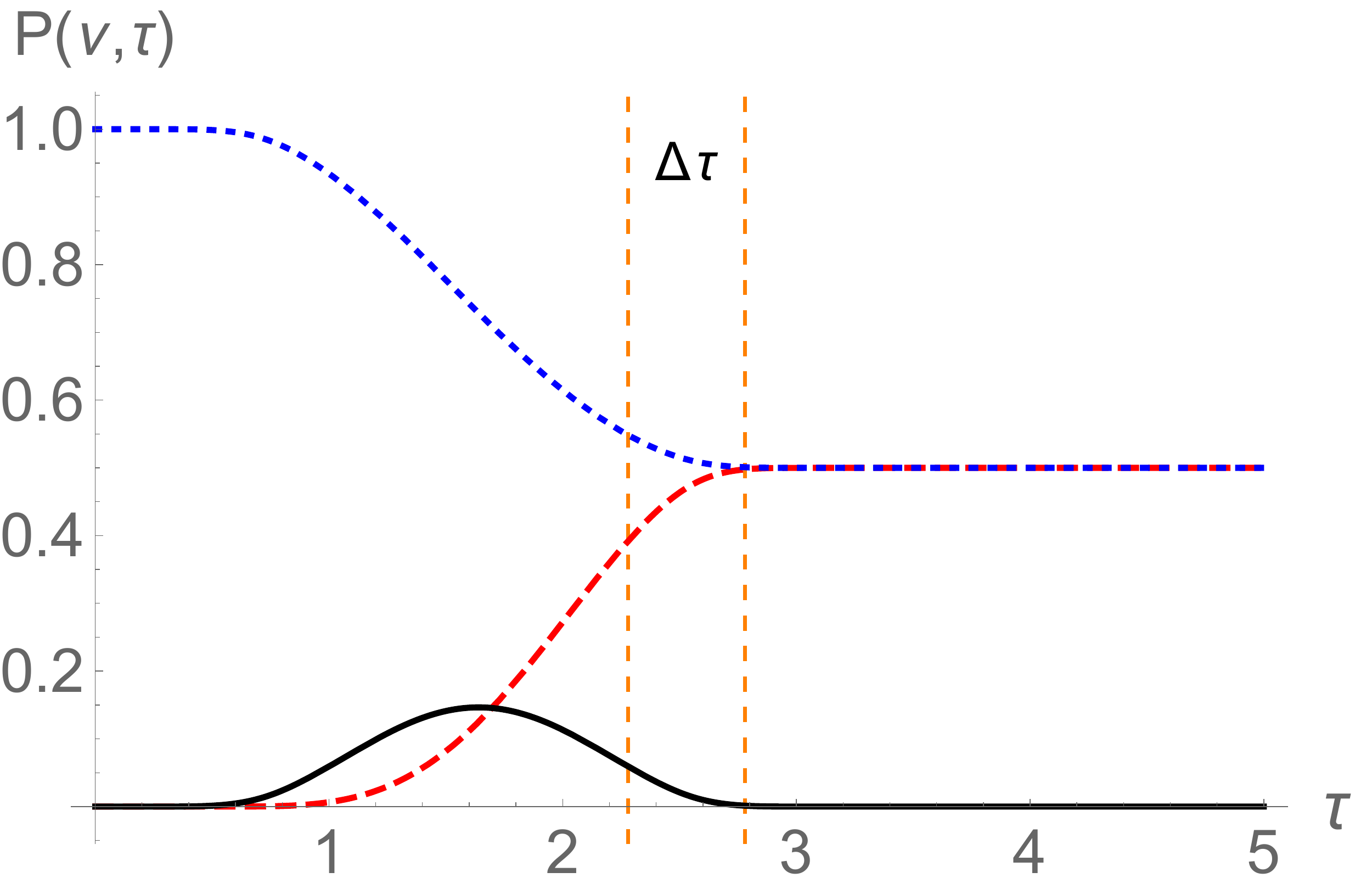}  \quad
		\includegraphics[width= 0.45\linewidth]{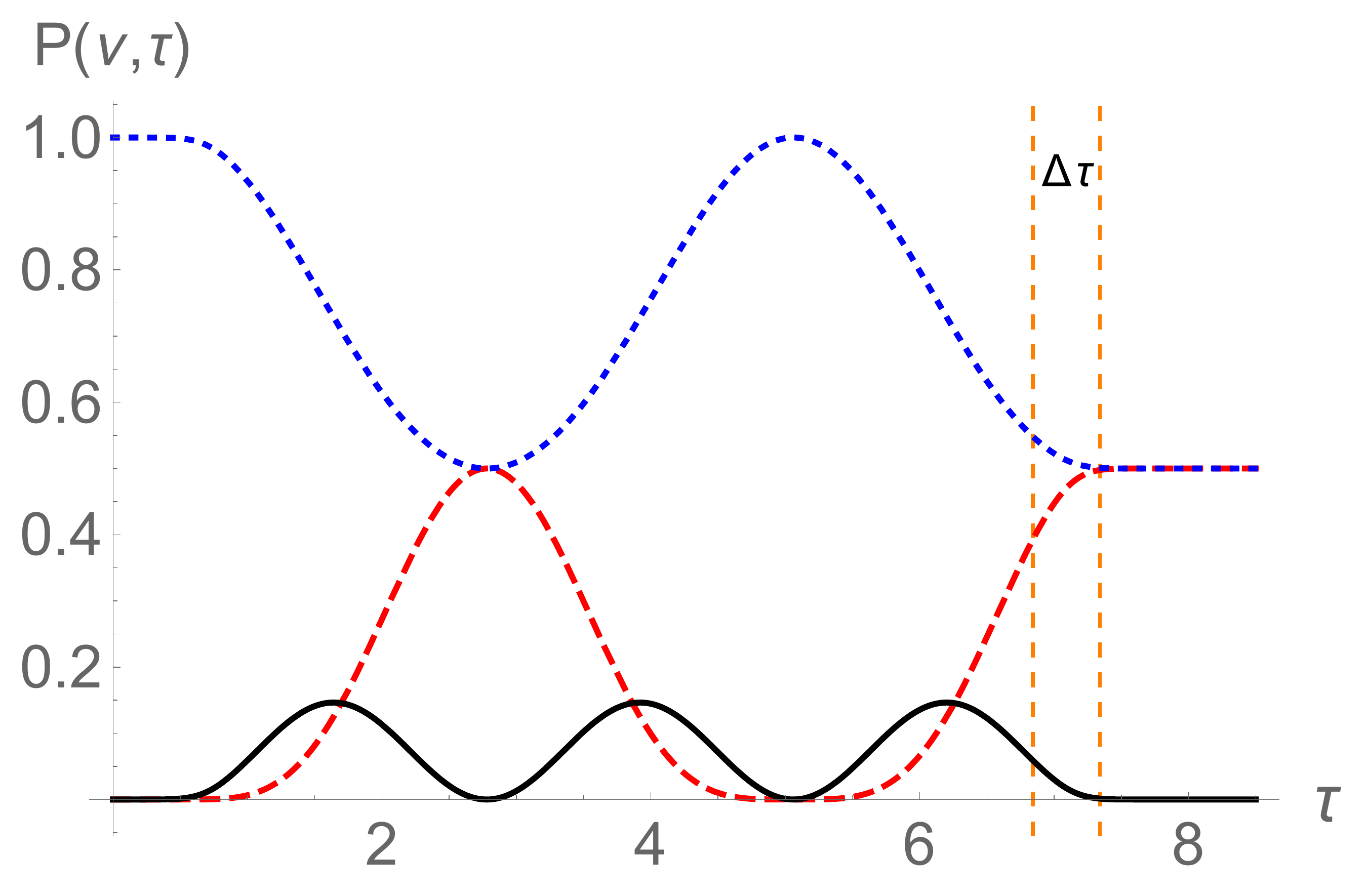} 
\end{center}
\caption{Photon number probability during the time inside the cavity for the first atom. Two different times have been chosen which annihilate the state component with $\nu-1$ photons (see text). The probabilities shown are ${\cal P}(\nu-2,\tau)$ (red, dashed line), ${\cal P}(\nu-1,\tau)$ (black, solid line), and ${\cal P}(\nu,\tau)$ (blue, dotted line). Here, $\Delta \tau = 1/2$ and the total excitation number is $M=\nu=3$.}
\label{ProbFot}
\end{figure}

\begin{figure}
\begin{center}
		\includegraphics[width=0.45\linewidth]{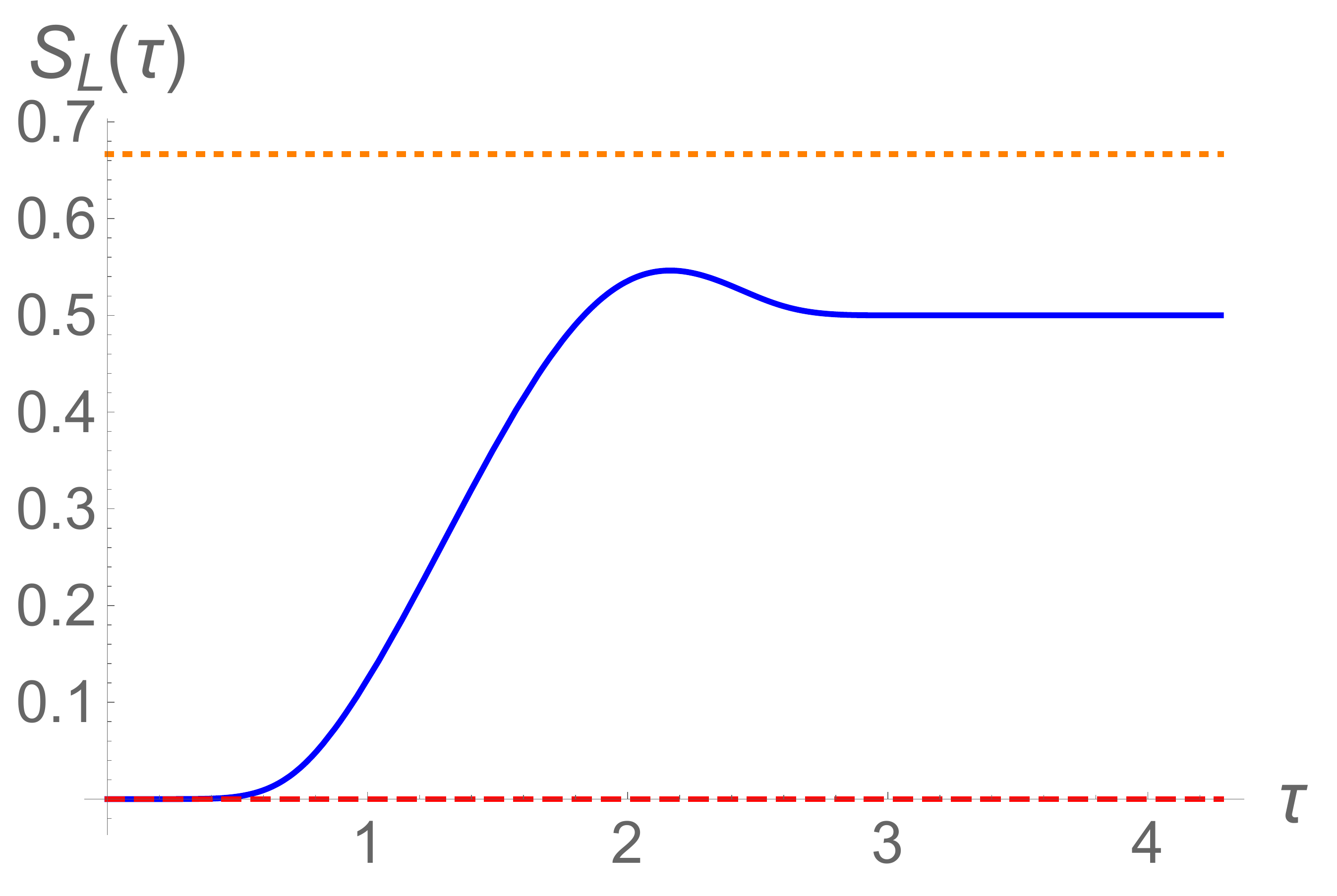}  \quad
		\includegraphics[width= 0.45\linewidth]{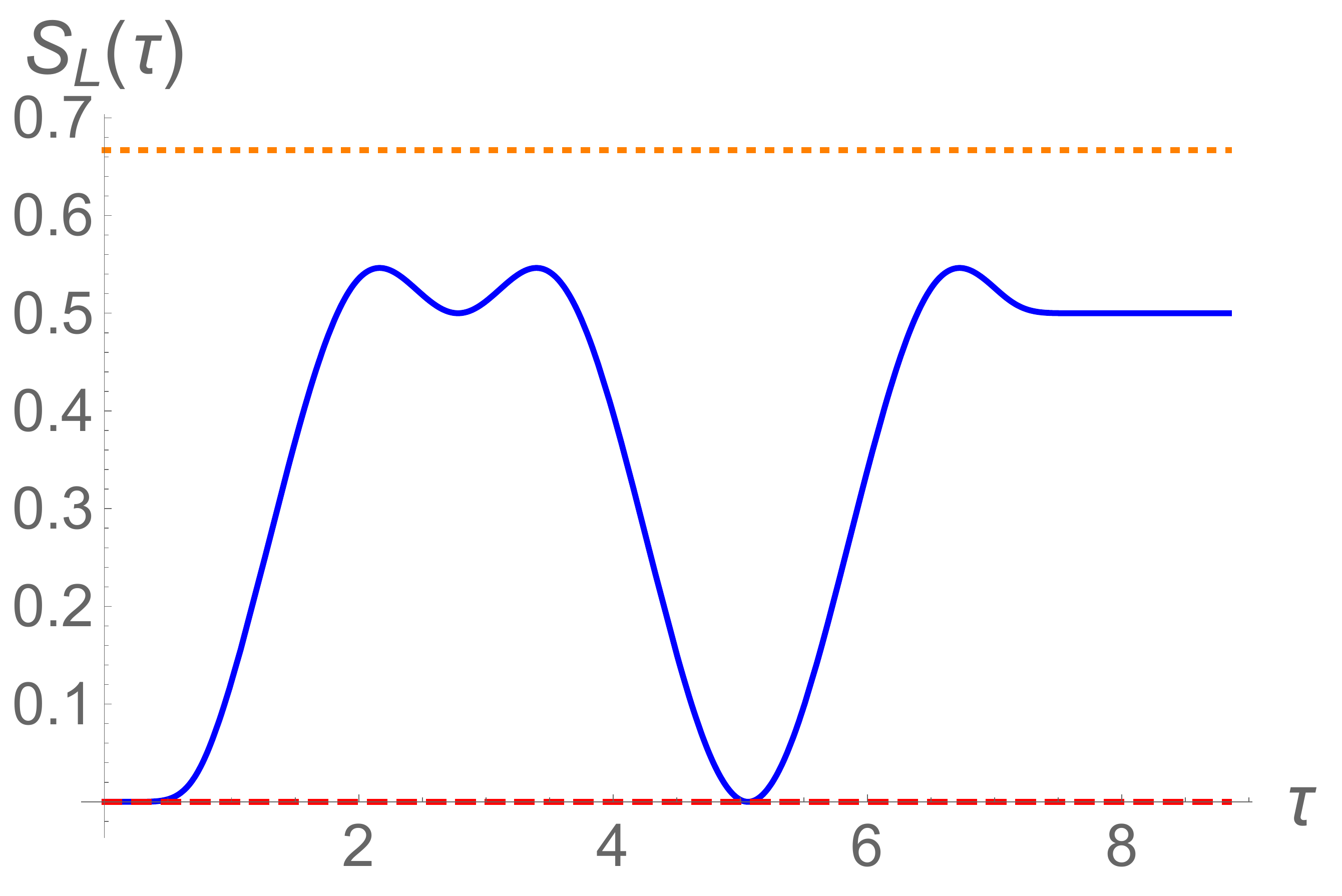} 
\end{center}
\caption{Evolution of the linear entropy inside the cavity for the first atom. Two different times have been chosen which annihilate the state component with $\nu-1$ photons: $\tau_1$ (left), and $\tau_2$ (right). For these figures we considered a total excitation number of $M=3$. The horizontal dashed lines correspond to the maximum and minimum possible values of the matter-field entanglement.}
\label{LinearEntropy}
\end{figure}

We now review in detail the procedure to generate dynamically the ${\cal C}_n$-state of light:

Firstly, we solve the system of differential equations~(\ref{sys.eqs}) for the first atom in the ground state coupled with the cavity with a definite number of photons $\vert \psi_1(0)\rangle= \vert \nu_0\rangle\otimes\vert 1\,1\,1\rangle$. Then we determine the properties of the system, for example: the atom-field correlations through the calculation of the von Neumann or linear entropy, the photon number probabilities, and the Husimi function for the electromagnetic field. The photon number probabilities during the time inside the cavity for the first atom are shown in Figure~\ref{ProbFot}. Two different exit times have been chosen which annihilate the state component with $\nu_0-1$ photons: $\tau_1 = t_s + \Delta \tau = \pi/{\cal E}_\Xi + \Delta \tau$ (left), and $\tau_2 = 3 \, t_s + \Delta \tau = 3\pi/{\cal E}_\Xi + \Delta \tau$ (right), where $t_s$ is given by Eq.~(\ref{times}). Recall that $t_s$ is the time at which the probability of finding $\nu=\nu_0-1$ photons vanishes, according to the analytic solution obtained in the previous section. When considering the coupling strengths dependent on time, there is however a delay equal to $\Delta \tau = 1/2$ to get the same condition. The different probabilities to find $\nu_0$, $\nu_0-1$, and $\nu_0-2$ photons, correspond to the ones shown in Fig.~\ref{pphoton}.  The linear entropy is shown for the same two different exit  times in Fig.~\ref{LinearEntropy}.

%Figura 3

%Figura 4

Thus, at $t_s + \Delta \tau$, the first atom through the cavity leaves the electromagnetic field in the superposition
\[
\cos\theta \vert \nu_0-2\rangle + e^{i \xi} \sin\theta \vert \nu_0\rangle \, .
\]

\begin{figure}[h!]
\begin{center}
		\includegraphics[width=0.43\linewidth]{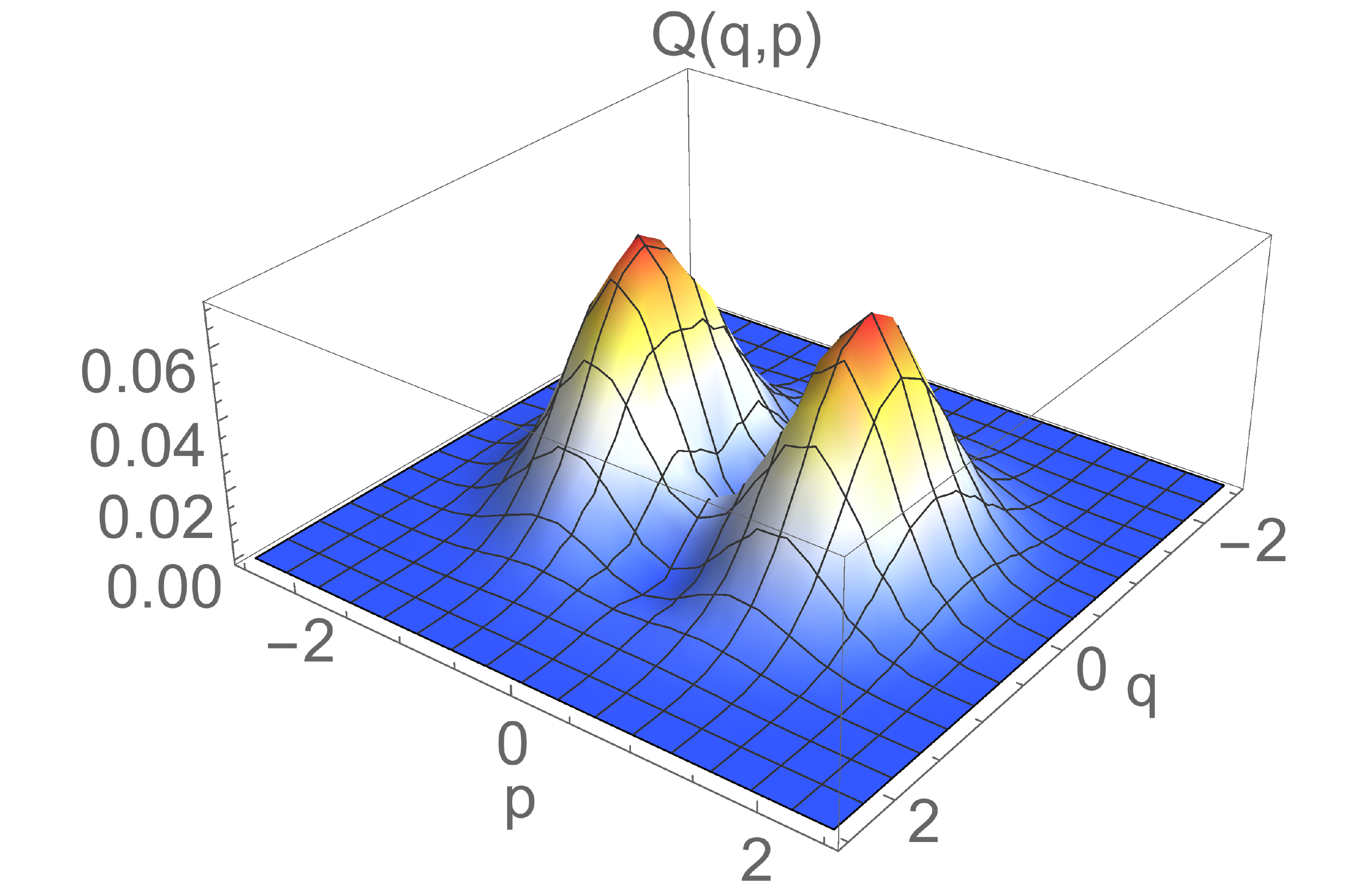} \quad
		\includegraphics[width=0.52\linewidth]{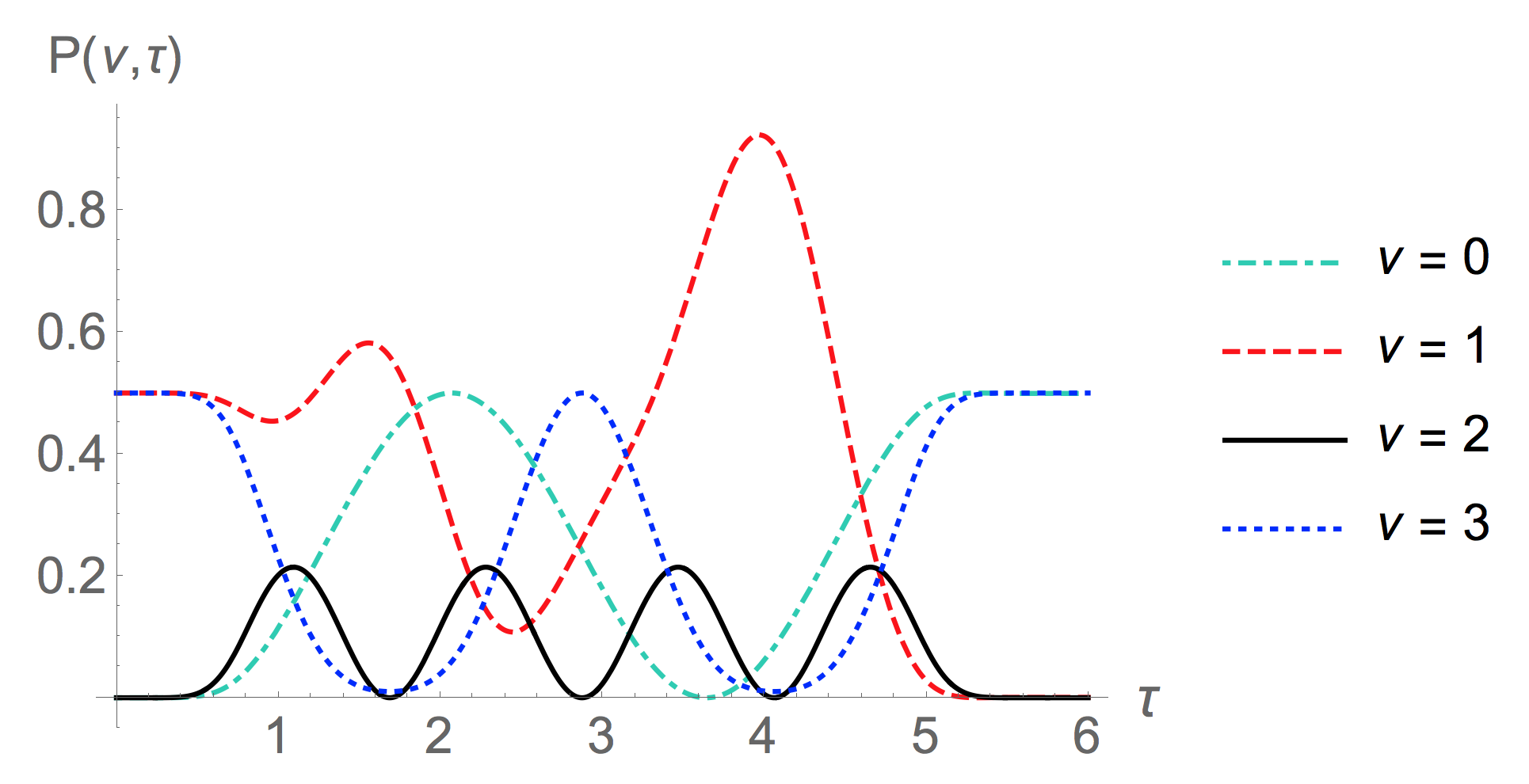}
\end{center}
\caption{Left: ${\cal C}_2$-state with $\nu$ and $\nu - 2$ photons, left by the passing of the first atom, and as seen by the second atom when it enters the cavity. In this case $\nu=3$. Right: Evolution of the photon number probability while the second atom traverses the cavity.}
\label{Qnu13}
\end{figure}

When a second atom enters the cavity we need to solve the system of differential equations~(\ref{sys.eqs}). Choosing $\nu_0=3$, at $\tau=0$ we have as the initial state the superposition given in equation~(\ref{init.state}), with $M_1=\nu_0 -2$ and $M_2=\nu_0$. By calculating the photon number probabilities for $\nu = 0,1,2,3$ we see that there is a tof time when the only surviving contributions are those for $\nu=0$ and $\nu=3$ photons [cf. Fig.~\ref{Qnu13} (right)]. The cavity is then left in a superposition with $\Delta M = 3$. If we want to study the dynamics of matter inside a cavity in a superposition of such states, we may follow the same procedure by sending a third atom through the cavity.

More generally, one may prepare in this manner a ${\cal C}_n$-state of light with an arbitrary difference in the total excitation number. If we start with $\nu_0$ photons inside the cavity, the first atom passing through will leave it in a superposition state of $\nu_0$ and $\nu_0 - 2$ photons, at an exit time $t_s + \Delta \tau$. A second atom entering the cavity will see a ${\cal C}_2$-light state as depicted in Fig.~\ref{Qnu13} (left), and will leave the cavity in a photon superposition of two Fock states as shown in Fig.~\ref{Qnu13} (right). While passing through the cavity it will allow us to obtain light states with various $\Delta\nu$ in superposition; an example is shown in Table~\ref{T1}, where $\mu_{12}(\tau) = \mu(\tau_{\rm tof},\,\tau),\ \mu_{23}(\tau) = \sqrt{2}\ \mu(\tau_{\rm tof},\,\tau)$. Note that in our construction $\Delta M = \Delta \nu$.

%Figura 5

%Table 1
\begin{table}[h!]
\caption{Photon number probabilities ${\cal P}(\nu,\,\tau)$ for the given tof time $\tau_{\rm tof}$, for the second atom in a cavity with a superposition of a total excitation number given by $M_1$ and $M_2$. When the atom leaves the cavity, the latter has a light superposition state with the $\Delta\nu$ shown. The probabilities are in sequential order for $\nu = \max\{0,\,M_1-2\}$ to $\nu = M_2$ photons.}
%\vspace{0.1in}
\label{T1}
\begin{center}
\begin{tabular}{c|c|c|c|c}
$M_1$ & $M_2$ & $\Delta\nu$ & $\tau_{\rm tof}$ & ${\cal P}(\nu,\,\tau)$\\
\hline&&&&\\[-6mm]
1 & 3 & 3 & 5.749 & {0.4993}, {0.0007}, {0.0000}, {0.5000} \\[2mm]
3 & 5 & 4 & 4.510 & {0.4851}, {0.0041}, {0.0108}, {0.0015}, {0.4985}\\[2mm]
1 & 5 & 5 & 2.685 & {0.4935}, {0.0065}, {0}, {0.0001}, {0.0082}, {0.4918}\\
\end{tabular}
\end{center}
\end{table}

\begin{figure}
\begin{center}
		\includegraphics[width=0.45\linewidth]{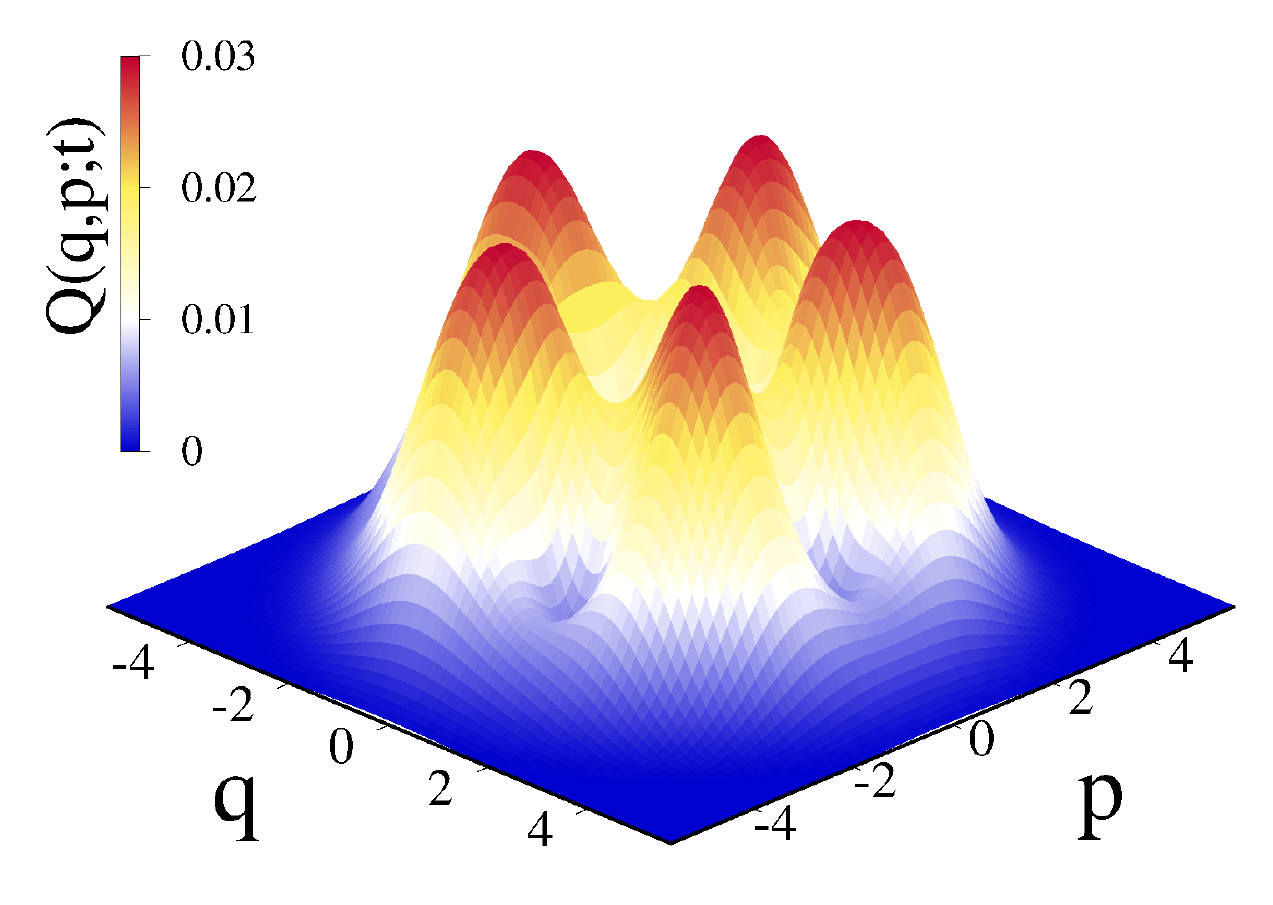}  \quad
		\includegraphics[width= 0.43\linewidth]{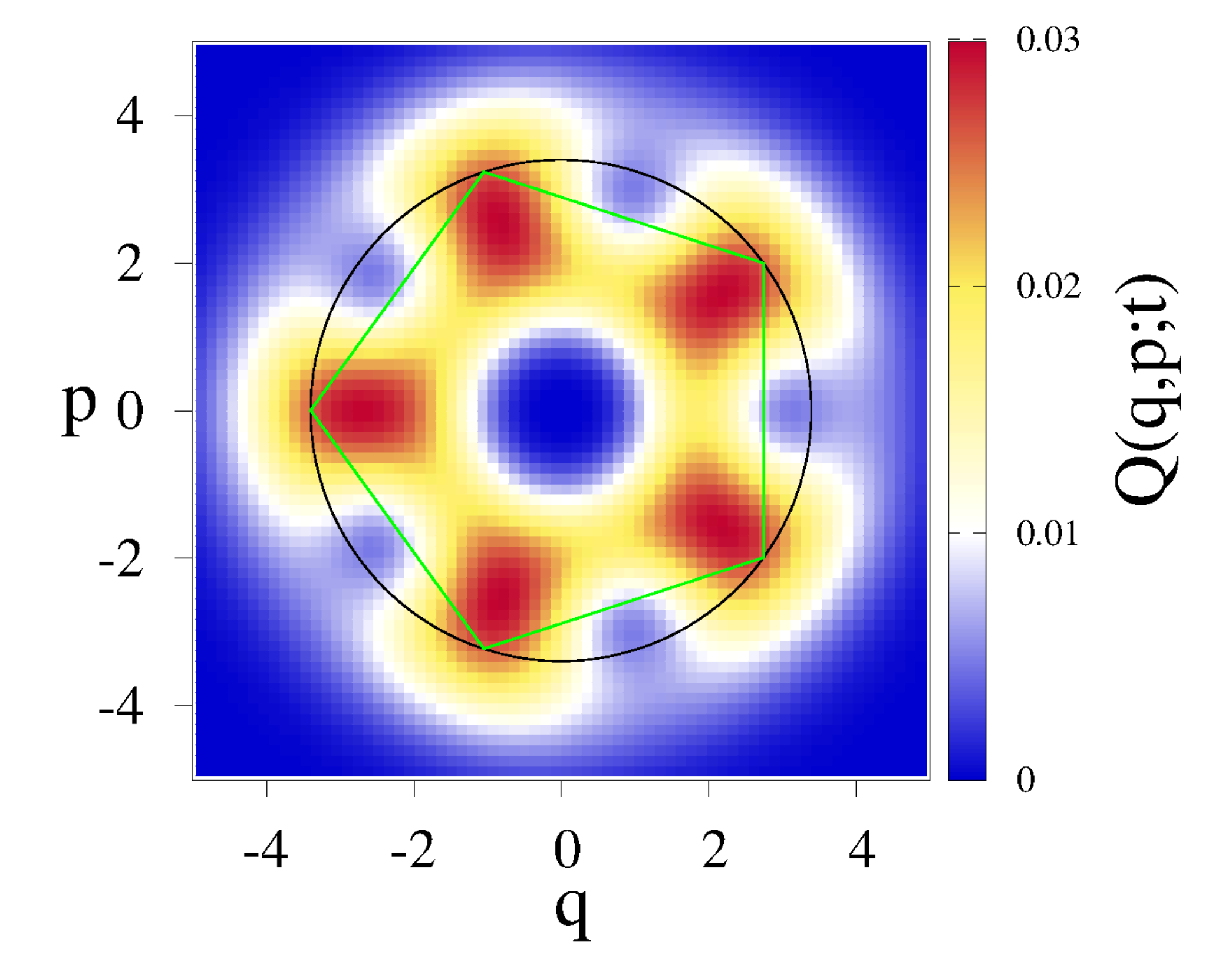} 
\end{center}
\caption{The complete Husimi function plotted as a function of the field quadratures, $\alpha=(q + i p)/{\sqrt 2}=\varrho\, e^{i \, \phi}$, for $\Delta M = \nu_2-\nu_1=5$. We use $\xi=0$, $\theta=\pi/4$, $\Delta_{12}=0$, and $\tau=\pi$, and we evaluate at $(\mu_{12},\,\mu_{23})=(1,\,\sqrt{2})$.}
\label{funQdet}
\end{figure}

For a $\Delta M = n$ the Husimi function given in eq.~(\ref{qh}) presents a cyclic point symmetry given by the group ${\cal C}_n$. The particular case $\Delta M = 5$ is depicted in Fig.~\ref{funQdet}. 

It is interesting to see the evolution of a light-superposition state in a cavity prepared as described above. An animation of the evolution, as given by the Husimi function, of a superposition of $2$ and $7$ photons with a $\Xi$-configuration atom in a cavity, is presented in the {\it Supplementary Material} for a time-step of $\delta\tau=\pi/32$. The effect of the parameter $\xi$ in the Husimi function itself, at a fixed time, is also shown as a rotation in the $q-p$ quadrature parameter space.

%Figura 6

\section{Extension to an arbitrary finite number of atoms}\label{sec5}

\begin{figure}
\begin{center}
		\includegraphics[width=0.45\linewidth]{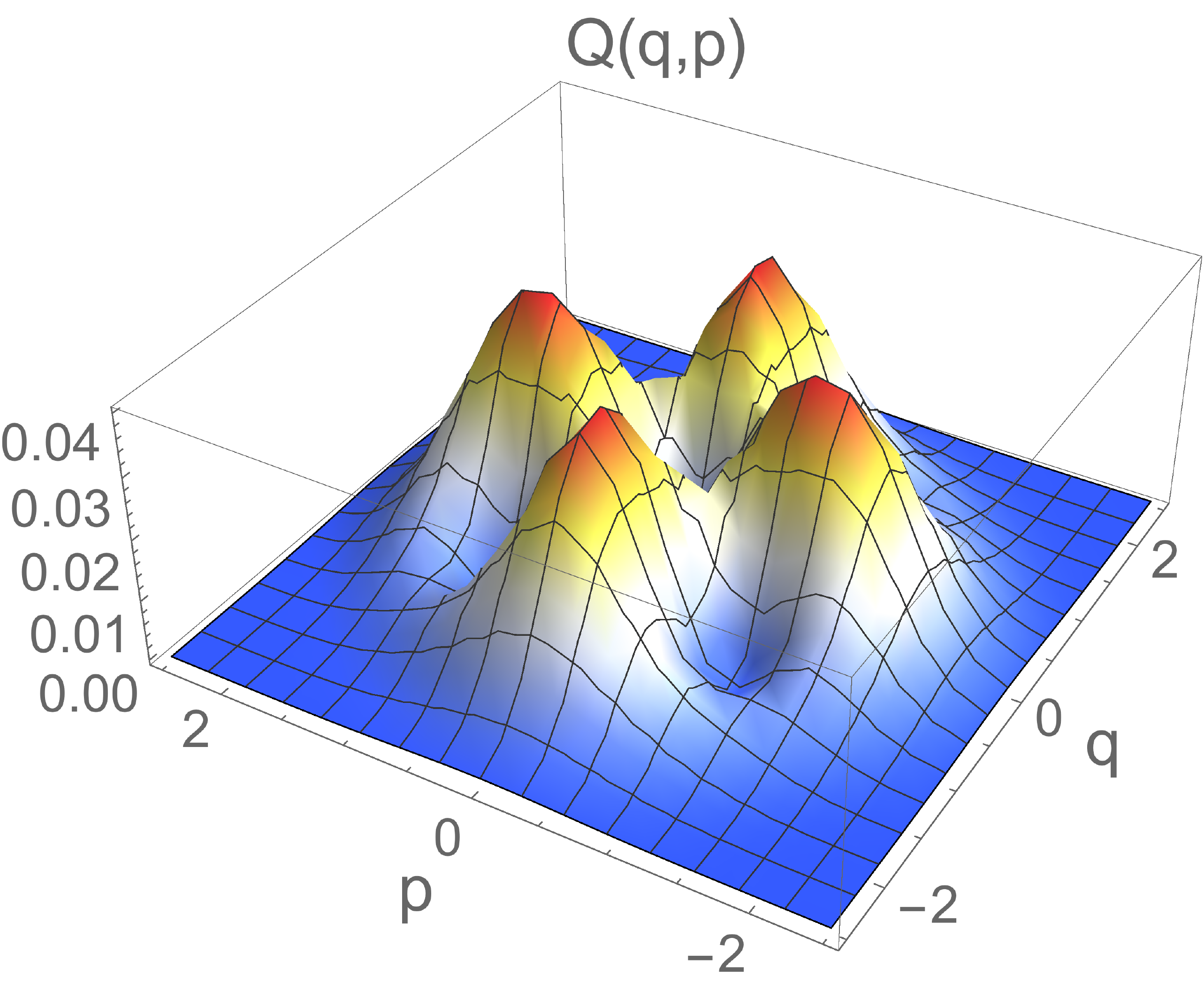}  \\[3mm]
		\includegraphics[width= 0.45\linewidth]{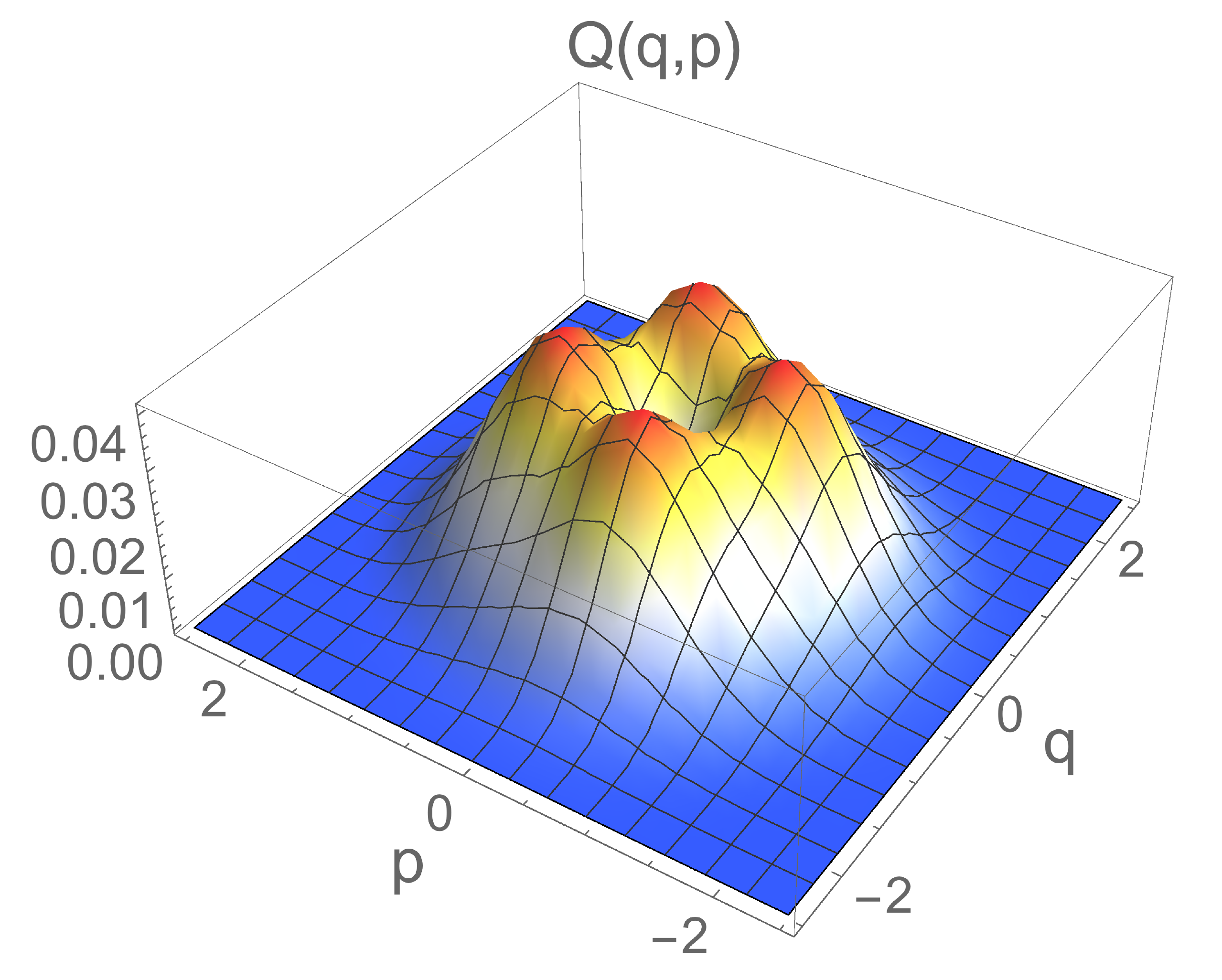} \\[3mm]
		\includegraphics[width= 0.45\linewidth]{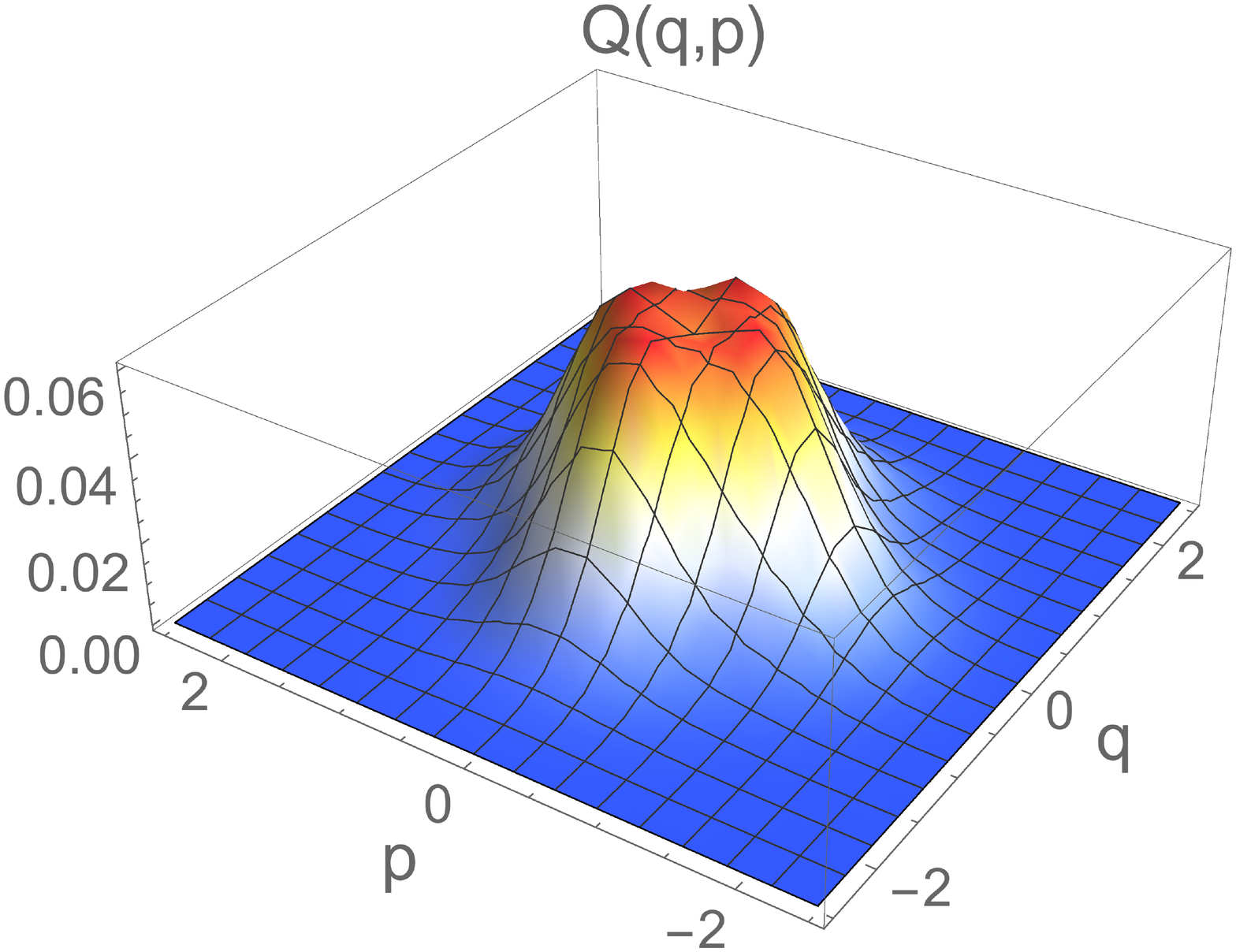}
\end{center}
\caption{The Husimi function plotted as a function of the field quadratures, for the evolution of two $3$-level atoms in their ground state having entered a cavity prepared in a ${\cal C}_4$-light state of $1$ and $5$ photons. We use $\xi=0$, $\theta=\pi/4$, $\Delta_{12}=0$, and $\tau=0$ (top left), $\tau=\tau_{\rm tof}/2$ (top right), and $\tau=\tau_{\rm tof}$ (bottom). We evaluate at $(\mu_{12},\,\mu_{23})=(1/\sqrt{2},\,1/\sqrt{2})$.}
\label{hq2atoms}
\end{figure}

The expressions (\ref{eq.psi.t}-\ref{sys.eqs}) can be generalised to any number of atoms $N_{a}$. In this case the dimension of the matrix system is given by the degeneracy of a three dimensional oscillator with $N_a$ quanta, because we consider $M \ge 2 \, N_a$. One may study the evolution of $N_a$ atoms inside a cavity prepared in a light superposition state as described in the last section. Figure~\ref{hq2atoms} shows this evolution for two $3$-level atoms in their ground state, in the $\Xi$-configuration, having entered a cavity prepared in a ${\cal C}_4$-light state of $1$ and $5$ photons, at $3$ different times: entering time $\tau=0$, half-way through $\tau=t_{tof}/2$, and leaving time $\tau=t_{tof}$. Similar plots are obtained at other values of $(\mu_{12},\,\mu_{23})$, for different numbers of atoms, and for all atomic configurations.

%Figura 7

Next we prove the invariance under ${\cal C}_n$ transformations of the Husimi function of light.
To that effect, we first establish the general solution of the Schr\"odinger equation as a time-dependent linear combination of the basis states introduced in Eq.~(\ref{base}),
\begin{equation}
\vert \Phi_M(\tau) \rangle = \sum_{q, r} C^{\textsc{m}}_{ q \, r} (t) \vert M-(q-r)\lambda_2-(N_a-q) \, \lambda_3; N_a\, q\, r\rangle \, .
\end{equation}
Since $M-(q-r)\lambda_2-(N_a-q) \, \lambda_3$ denotes the number of photons, we will write it as $\nu (M,q,r)$. 
The corresponding density matrix is constructed, then one takes the partial trace with respect to the matter components to get the reduced density matrix of the electromagnetic field, and calculate its expectation value with respect to the coherent state of light to find the Husimi function
\begin{equation}
\hspace{2.5 cm}
Q^{\textsc{m}}_H(\alpha, t) =  \sum_{q, r} \vert C^{\textsc{m}}_{ M\,q \, r} (t) \vert^2  \, \frac{e^{-\vert \alpha\vert^2}}{\nu (M,q,r)!} \, \vert \alpha \vert^{2 \, \nu (M,q,r)} \, .
\end{equation}
It is immediate that this Husimi function is invariant under rotations because it is a function of $\alpha=\rho \, e^{i \, \phi}$. 
We now wish to show that a linear superposition of two or more Fock states leads to a Husimi function with a discrete symmetry. To that end we consider the initial state
\begin{eqnarray}
\label{init.state2}
\vert\psi(0)\rangle &=& \left(\cos\theta \vert \nu_1\rangle + e^{i \xi} \sin\theta \vert \nu_2\rangle\right)\otimes\,
\vert N_a\,N_a\,N_a\rangle \, , \nonumber \\
&=& \cos\theta \, \vert \Phi_{\nu_1}(0) \rangle + e^{i \xi} \sin\theta \, \vert \Phi_{\nu_2}(0) \rangle \, ,
\end{eqnarray}
This state has the symmetry of $\mathcal{C}_{\nu_2 - \nu_1}$. As the cavity evolves in the presence of the $N_a$ atoms a complete superposition of many Fock states of light will arise (together with their corresponding atomic excitations), all of which obey the $\mathcal{C}_{\nu_2 - \nu_1}$ symmetry since the Hamiltonian preserves this symmetry at all times.
By following the same procedure as for the one-particle case the Husimi function takes the form
\begin{eqnarray}
Q_H(\nu_1,\nu_2,\alpha,\theta,\xi,\tau)&=& \cos^2\theta \, Q^{\nu_1}_H(\alpha, t) + \sin^2\theta \, Q^{\nu_2}_H(\alpha, t) \nonumber \\
&+ & \frac{1}{2} \sin2\theta \, \left(e^{-i \xi} {\cal F}_{\nu_1, \nu_2}(\alpha,t) + c.c. \right)\, ,
\end{eqnarray}
where we have defined 
\[
{\cal F}_{\nu_1, \nu_2}(\alpha,t)= \sum_{q, r}  C^{\nu_1}_{ q \, r} (t) \, C^{*}{}^{\nu_2}_{ q \, r} (t) \, \frac{e^{-\vert \alpha\vert^2} \, \rho^{\nu_1+\nu_2 -2(q-r)\lambda_2 - 2(N_a-q)\lambda_3} \, e^{i \phi (\nu_1-\nu_2)}}{\sqrt{\nu (\nu_1,q,r)! \, \nu (\nu_2,q,r)!}} \, .
\]
so that $Q_H$ is invariant under the transformation $\phi \to \phi + 2\pi/\vert\nu_1-\nu_2\vert$, as we wanted to prove. We have thus left the cavity with an electromagnetic field described by a superposition of Fock states with a $\mathcal{C}_{\vert\nu_1 - \nu_2\vert}$ symmetry. For the case $\nu_1=2,\,\nu_2=7$, it has contributions from $\nu=0,1,\ldots,7$ components.

\section{Conclusions}\label{sec6}

In this work we have shown how to generate superpositions of photon number operator states within the generalised Tavis-Cummings model (GTC), independently of the atomic dipolar strengths and of the number of atoms.

The procedure for a possible experimental setup was presented to generate these state superpositions with a fixed difference of the total excitation number, $\Delta M=|M_1-M_2|$, and it was shown that the field sector is invariant under point transformations of ${\cal C}_n$, the cyclic group in $n$ dimensions, with $n=|M_1-M_2|$.

The stationary states of the GTC model, for the one-particle case, were given in analytic form together with the corresponding evolution operator. This operator was used to study the dynamics of an arbitrary initial state. By appropriately selecting the time-of-flight of an atom through a resonant cavity, we show how to obtain crystallised states of light. In particular, we considered a superposition of two states with $M_1$ and $M_2$ total excitation numbers, and the crystallised states were given explicitly.  The results for the one-atom case were extended to any number of atoms, both under resonant conditions and with detuning.

In the {\it Supplementary Material} we exhibit: the dynamics of the Husimi function associated to a single atom inside a cavity with $\nu_1=2$ and/or $\nu_2=7$ photons, the rotation effect of the parameter $\xi$ of the initial state~(\ref{init.state}), and the corresponding symmetry associated to the cyclic group of dimension $\nu_2-\nu_1$. (See Appendix~\ref{app3} for details.)

\section*{Acknowledgements}
This work was partially supported by CONACyT-M\'exico (under Project No.~238494), and DGAPA-UNAM (under Project No.~IN101217).

\appendix
\section{Dressed states for a $3$-level atom}\label{app1}
The dressed states for a single atom can be obtained in analytic form. They are explicitly given for each atomic configuration in what follows: 

\smallskip

\noindent
For the $\Xi$-configuration with the detuning conditions $\Delta_{12} + \Delta_{23}=0$: 
\begin{eqnarray}
&& 
\vert \psi_0\rangle_\Xi =  -\frac{\sqrt{M_\Xi} \, \mu_{12}}{{\Omega}_\Xi}  \vert M_{\Xi}-2; 1\, 0 \, 0\rangle + \frac{\sqrt{M_\Xi-1} \, \mu_{23}}{{\Omega}_\Xi}  \, \vert M_\Xi; 1\, 1 \, 1\rangle \, , \nonumber \\[1mm]
&& \vert \psi_\pm\rangle_\Xi = \frac{1}{{\cal E}_\Xi \left( 2 \pm \frac{\Delta_{12}}{{\cal E}_\Xi}\right)^{1/2}} \left\{\sqrt{M_\Xi-1} \, \mu_{23} \vert M_{\Xi}-2; 1 \,  0 \, 0\rangle  + \sqrt{M_\Xi} \, \mu_{12} \, \vert M_\Xi; 1 \, 1 \, 1 \rangle  \right. \nonumber \\
&& \left. \hspace{1.5cm} - \Bigl(\Delta_{12}/{2} \pm {\cal E}_\Xi \Bigr) \, \vert M_\Xi-1; 1 \, 1 \, 0 \rangle \right\} \, . 
\end{eqnarray}

\smallskip

\noindent
For the $\Lambda$-configuration with $\Delta_{13} - \Delta_{23}=0$:
\begin{eqnarray}
&&\vert \psi_0\rangle_\Lambda =   -\frac{\mu_{13}}{\sqrt{\mu^2_{13} + \mu^2_{23}}} \, \vert M_{\Lambda}; 1 \, 1 \, 0\rangle + \frac{\mu_{23}}{\sqrt{\mu^2_{13} + \mu^2_{23}}}  \, \vert M_\Lambda; 1 \, 1 \,  1\rangle  \, , 
\nonumber  \\[2mm]
&&  \vert \psi_\pm\rangle_\Lambda =   \frac{1}{{\cal E}_\Lambda \Bigl( 2 \pm \frac{\Delta_{13}}{{\cal E}_\Lambda}\Bigr)^{1/2}} \left\{ \sqrt{M_\Lambda} \, \mu_{13}  \, \vert M_{\Lambda}; 1 \, 1 \,  1\rangle - \Bigl(\pm \frac{\Delta_{13}}{2} + {\cal E}_\Lambda \Bigr)  \vert M_{\Lambda}-1; 1 \, 0 \, 0\rangle \right.
\nonumber \\
&& \qquad \quad \left. \quad + \sqrt{M_\Lambda} \, \mu_{23}  \, \vert M_\Lambda; 1 \,  1\,  0 \rangle  \right\}   \, . 
\end{eqnarray}

\smallskip

\noindent
For the $V$-configuration with $\Delta_{12} - \Delta_{13}=0$:
\begin{eqnarray}
&& \vert \psi_0\rangle_V =   -\frac{\mu_{12}}{\sqrt{\mu^2_{12} + \mu^2_{13}}} \, \vert M_{V}-1; 1 \,  0 \, 0\rangle + \frac{\mu_{13}}{\sqrt{\mu^2_{12} + \mu^2_{13}}}  \, \vert M_V-1; 1\, 1\, 0 \rangle  \, ,  \nonumber 
\\[1mm]
&& \vert \psi_\pm\rangle_V =  \frac{1}{{\cal E}_V \Bigl( 2 \mp \frac{\Delta_{12}}{{\cal E}_V}\Bigr)^{1/2}} \left\{\mp \sqrt{M_V} \, \mu_{13}  \vert M_{V}-1; 1 \, 0 \, 0 \rangle \mp \sqrt{M_V} \, \mu_{12} \, \vert M_V-1; 1 \, 1\, 0 \rangle \right. \nonumber  \\
&& \left. \hspace{1.7cm} + \Bigl( \mp \frac{\Delta_{12}}{2} + {\cal E}_V \Bigl) \, \vert M_V; 1 \, 1 \, 1\rangle \right\} \, .\end{eqnarray}

\section{Evolution Operator for a $3$-level atom}
\label{app2}

This operator, in the interaction picture, can be determined in analytic form for the detuning conditions indicated in~\ref{app1}. They take the following expressions for each atomic configuration:

\smallskip

\noindent
For the $\Xi$-configuration:
\begin{eqnarray}
&& U_I(\tau)_{11} = \frac{1}{{\cal E}^2_\Xi - \Delta^2_{12}/4}
\left\{ M_\Xi \, \mu^2_{12} + (M_\Xi-1) \, \mu^2_{23} \Bigl( \cos{{\cal E}_\Xi \tau} + i \Delta_{12} \frac{\sin{{\cal E}_\Xi \tau}}{{2\, \cal E}_\Xi} \Bigr) e^{-i \Delta_{12} \, \tau/2} \right\} \, , \nonumber \\[2mm]
&&  U_I(\tau)_{12}= i \sqrt{M_\Xi-1} \, \mu_{23} \frac{\sin{{\cal E}_\Xi \tau}}{{\cal E}_\Xi} \, e^{-i \Delta_{12} \, \tau/2} \, ,  \nonumber
\\[2mm] 
&& U_I(\tau)_{13} = \frac{\sqrt{M_\Xi (M_\Xi-1) } \, \mu_{12} \, \mu_{23}}{{\cal E}^2_\Xi - \Delta^2_{12}/4} 
 \left\{ -1 + \Bigl(\cos{{\cal E}_\Xi \tau} + i \Delta_{12} \frac{\sin{{\cal E}_\Xi \tau}}{{2 \, \cal E}_\Xi} \Bigr) e^{-i \Delta_{12} \, \tau/2} \right\} \, ,   \nonumber \\[2mm]
&&  U_I(\tau)_{22} =  \Bigl(\cos{{\cal E}_\Xi \tau} - i \Delta_{12} \frac{\sin{{\cal E}_\Xi \tau}}{ 2\,{\cal E}_\Xi} \Bigr) e^{-i \Delta_{12} \, \tau/2} \, ,   \\[2mm]
&&U_I(\tau)_{23}= i \sqrt{M_\Xi} \ \mu_{12} \frac{\sin{{\cal E}_\Xi \tau}}{{\cal E}_\Xi} \, e^{-i \Delta_{12} \, \tau/2} \, ,  \nonumber
\\[2mm] 
&& U_I(\tau)_{33} = \frac{1}{{\cal E}^2_\Xi - \Delta^2_{12}/4}
\left\{ (M_\Xi-1)\, \mu^2_{23} +  M_\Xi \, \mu^2_{12} \Bigl( \cos{{\cal E}_\Xi \tau} + i \Delta_{12} \frac{\sin{{\cal E}_\Xi \tau}}{{2\, \cal E}_\Xi} \Bigr) e^{-i \Delta_{12} \, \tau/2} \right\} \, .  \nonumber
\end{eqnarray}

\smallskip

\noindent
For the $\Lambda$-configuration:
\begin{eqnarray}
&& U_I(\tau)_{11} = e^{-i \Delta_{13} \, \tau/2}
\left\{ \cos{{\cal E}_\Lambda \tau} - i \, \Delta_{13} \frac{\sin{{\cal E}_\Lambda \tau}}{{2\, \cal E}_\Lambda}  \right\} \, , \nonumber \\
&& U_I(\tau)_{12} = \frac{ -i \, \sqrt{M_\Lambda}\,  \mu_{23}\, e^{-i \Delta_{13} \, \tau/2}}{{\cal E}_\Lambda} \, \sin{{\cal E}_\Lambda \tau}  \, , \nonumber \\[2mm]
&& U_I(\tau)_{13} = \frac{- i \, \sqrt{M_\Lambda}\,  \mu_{13}\, e^{-i \Delta_{13} \, \tau/2}}{{\cal E}_\Lambda}
  \, \sin{{\cal E}_\Lambda \tau}  \, , \nonumber \\[2mm]
&& U_I(\tau)_{22} = \frac{M_\Lambda\, }{{\cal E}^2_\Lambda - \Delta^2_{13}/4}
\left\{ \mu^2_{13} + \mu^2_{23} e^{-i \Delta_{13} \, \tau/2}  \Bigl( \cos{{\cal E}_\Lambda \tau} + i \, \Delta_{13} \frac{\sin{{\cal E}_\Lambda \tau}}{{2\, \cal E}_\Lambda} \Bigr) \right\} \, , \\[2mm]
&& U_I(\tau)_{23} = -\frac{M_\Lambda\,  \mu_{12}\, \mu_{23} }{{\cal E}^2_\Lambda - \Delta^2_{13}/4}
\left\{ 1 - e^{-i \Delta_{13} \, \tau/2} \, \Bigl(\cos{{\cal E}_\Lambda \tau} + i \, \Delta_{13} \frac{\sin{{\cal E}_\Lambda \tau}}{{2\, \cal E}_\Lambda} \Bigr) \right\} \, , \nonumber \\[2mm]
&& U_I(\tau)_{33} = \frac{M_\Lambda\, }{{\cal E}^2_\Lambda - \Delta^2_{13}/4}
\left\{ \mu^2_{23}  +  \mu^2_{13}  e^{-i \Delta_{13} \, \tau/2} \Bigl( \cos{{\cal E}_\Lambda \tau} + i \Delta_{13} \frac{\sin{{\cal E}_\Lambda \tau}}{{2\, \cal E}_\Lambda} \Bigr) \right\} \, . \nonumber 
\end{eqnarray}

\smallskip

\noindent
For the $V$-configuration:
\begin{eqnarray}
&& U_I(\tau)_{11} = \frac{M_V\, }{{\cal E}^2_V - \Delta^2_{12}/4}
\left\{ \mu^2_{12} e^{-i \Delta_{12} \, \tau} + \mu^2_{13} e^{-i \Delta_{12} \, \tau/2}  \Bigl( \cos{{\cal E}_V \tau} - i \, \Delta_{12} \frac{\sin{{\cal E}_V \tau}}{{2\, \cal E}_V} \Bigr) \right\} \, , \nonumber \\[2mm]
&& U_I(\tau)_{12} = -\frac{M_V\,  \mu_{12}\, \mu_{13} }{{\cal E}^2_V - \Delta^2_{12}/4}
\left\{ e^{-i \Delta_{12} \, \tau} - e^{-i \Delta_{12} \, \tau/2} \, \Bigl(\cos{{\cal E}_V \tau} - i \, \Delta_{12} \frac{\sin{{\cal E}_V \tau}}{{2\, \cal E}_V} \Bigr) \right\} \, , \nonumber \\[2mm]
&& U_I(\tau)_{13} = \frac{ i \, \sqrt{M_V}\,  \mu_{13}\, e^{-i \Delta_{12} \, \tau/2}}{{\cal E}_V}
  \, \sin{{\cal E}_V \tau}  \, ,  \\[2mm]
&& U_I(\tau)_{22} = \frac{M_V\, }{{\cal E}^2_V - \Delta^2_{12}/4}
\left\{ \mu^2_{13} e^{-i \Delta_{12} \, \tau} +  \mu^2_{12}  e^{-i \Delta_{12} \, \tau/2} \Bigl( \cos{{\cal E}_V \tau} - i \Delta_{12} \frac{\sin{{\cal E}_V \tau}}{{2\, \cal E}_V} \Bigr) \right\} \, , \nonumber \\[2mm]
&& U_I(\tau)_{23} = \frac{ i \, \sqrt{M_V}\,  \mu_{12}\, e^{-i \Delta_{12} \, \tau/2}}{{\cal E}_V}
  \, \sin{{\cal E}_V \tau}  \, , \nonumber \\[2mm]
&& U_I(\tau)_{33} = e^{-i \Delta_{12} \, \tau/2}
\left\{ \cos{{\cal E}_V \tau} + i \, \Delta_{12} \frac{\sin{{\cal E}_V \tau}}{{2\, \cal E}_V}  \right\} \, , \nonumber
\end{eqnarray}
and for each configuration the corresponding hermitian conjugates.

 Note that for the $\Lambda$- and $V$-configurations the dependence in the total number of excitations appears only in the argument of the trigonometric functions.

\section{Evolution of the Husimi function of light}\label{app3}

The corresponding animations of the material presented here appear in the {\it Supplementary Material}.

\subsection{QED cavity with 2 or 7 photons}

As an example we consider $3$-level atoms in the $\Xi$-configuration with the following parameters: $\Omega=1$ for the field frequency,  $\omega_1=0,\,\omega_2=1,\,\omega_3=2$ for the atomic level frequencies ($\hbar=1$), and dipolar strengths $\mu_{12}=1$ and $\mu_{23}=\sqrt{2}$.

The most general superposition of two states with different values of the total number of excitations is given by
\begin{equation}\label{eq.psi0.1}
|\psi_0\ket =\cos(\theta)\,|\nu_1;N_a\,\,q_1\,r_1\ket +e^{i\xi}\,\sin(\theta) \,|\nu_2;\,N_a\,q_2\,r_2\ket,
\end{equation}
where each state is given by the direct product of the photon (Fock basis) and matter (Gelfand basis) states $|\nu;\,N_a\,q\,r\ket = |\nu\ket\otimes|N_a\,q\,r\ket$. In terms of these quantum numbers, the total number of excitations for $N_a$ atoms in the $\Xi$-configuration reads as $M=\nu +2N_a -q-r$.

For the numerical calculation we consider the particular case of a single particle $N_a=1$ in its ground state, i.e., $q_1=q_2=r_1=r_2=N_a$, and the cavity in a superposition of states with $\nu_1=2$ and $\nu_2=7$ photons. For this case the initial state is separable and takes the form
\begin{equation}\label{eq.psi0.2}
|\psi_0\ket =\left[\cos(\theta)\,|\nu_1\ket +e^{i\xi}\,\sin(\theta) \,|\nu_2\ket\right]\otimes |N_a\,N_a\,N_a\ket.
\end{equation}

%
%\begin{center}
%\begin{figure}[t!]
%\begin{center}
%\animategraphics[width = 0.45\linewidth,controls,buttonsize=1.8em,loop]{5}{Qast_theta0/xiast_theta0}{0}{64}\hspace{0.05\linewidth}
%\animategraphics[width = 0.45\linewidth,controls,buttonsize=1.8em,loop]{5}{Qast_thetapi2/xiast_thetap2}{0}{64}
%\end{center}
%\caption{State given by eq.~(\ref{eq.psi0.2}) with $\xi=0$, $\theta=0$ (left) and $\theta=\pi/2$ (right). The Husimi quasi-probability is shown as a function of dimensionless time $\tau$. The animation is presented in steps $\delta\tau=\pi/32$. See main text.
%}\label{f.Qast0}
%\end{figure}
%\end{center}
%

In Animations 1a, 1b in the Supplementary Material the Husimi quasi-probability is shown as a function of dimensionless time $\tau=\Omega\,t$ for the case of an initial state with $\xi=0$ and $\theta=0$ (left), which has a single $M=\nu_1$ contribution; and for an initial state with $\xi=0$ and $\theta=\pi/2$ (right), which corresponds to $M=2$ and $M=7$ excitations. In the animations one may observe that both present a radial symmetry and that their profiles evolves as volcano shapes. The radius of the volcano mouth oscillates in the range $\sqrt{M-2} \leq R \leq \sqrt{M}$ as a function of time. $R$ is defined as the square root of the expectation value of the number of photons, $R^2 = (1/2)(q^2 + p^2)$. Thus for the state with $M=2$ the volcano shape can evolve into a single peak, which corresponds to the vacuum state of the cavity.

\subsection{${\cal C}_5$ light state}
Animation 2 in the Supplementary Material describes the evolution of a $3$-level atom in the $\Xi$-configuration inside a cavity, in a superposition of photon states as in Eq.~(\ref{eq.psi0.2}) with $\nu_1=2$ and $\nu_2=7$.

%%
%\begin{center}
%\begin{figure}[h!]
%\begin{center}
%\animategraphics[width = 0.8\linewidth,controls,buttonsize=3em,loop]{5}{Qast3D/xiast_}{0}{64}
%\end{center}
%\caption{The Husimi quasi-probability as a function of the dimensionless time $\tau$ is shown for the state~(\ref{eq.psi0.2}) with  $\xi=0$ and $\theta=\pi/4$. The animation is presented in steps $\delta\tau=\pi/32$. See main text.}\label{f.Qast}
%\end{figure}
%\end{center}

\subsection{Effect of the parameter $\xi$}
The effect of the $\xi$-parameter in the Husimi function is shown in Animation 3, in the Supplementary Material. The initial state is taken to be that considered above in the ${\cal C}_5$-case. We take $\theta=\pi/4$, $\tau=\pi/2$, and analyse the contour plot of the Husimi function as function of $\xi$. 

%
%\begin{center}
%\begin{figure}[h!]
%\begin{center}
%\animategraphics[width = 0.8\linewidth,controls,buttonsize=3em,loop]{5}{Qasxi/xi_}{0}{32}
%\end{center}
%\caption{The Husimi quasi-probability at dimensionless time $\tau=\pi/2$ as a function of the $\xi$-parameter is shown for the state~(\ref{eq.psi0.2}) with $\theta=\pi/4$. The animation is presented in steps $\delta\xi=\pi/16$. See main text.}
%\label{f.Qasxi}
%\end{figure}
%\end{center}
%

%

\end{document}